\documentclass[pdflatex, sn-mathphys]{sn-jnl}

\usepackage{graphicx}%
\usepackage{multirow}%
\usepackage{amsmath,amssymb,amsfonts}%
\usepackage[title]{appendix}
\usepackage{xcolor}
\usepackage{textcomp}

\usepackage{ulem}
\usepackage{lmodern}
\usepackage{aas_macros}
\usepackage{color}
\usepackage{hyperref}
\usepackage{multibib} 
\usepackage{placeins}

\usepackage{float}

\usepackage{xcolor}

\usepackage{mathrsfs}

\usepackage{placeins}
\usepackage{float}
\usepackage{nameref}

\usepackage[markup=default]{changes}
\setauthormarkup{\relax}             
\definechangesauthor[color=black]{anon}
\setauthormarkup{} 
\definechangesauthor[color=black]{anon2}
\setauthormarkup{} 
\definechangesauthor[color=black]{anon3}

\newcites{Methods}{Methods References}

\usepackage{hyperref}
\usepackage{cleveref}

\newcommand{\EDref}[1]{Extended Data Fig.~\ref{#1}}

\begin{document}

\title[Article Title]{The nature of the Norma arm as a bar-driven star-forming shock front}

\author*[1]{\fnm{Hillary} \sur{Davis}}\email{hdav7324@uni.sydney.edu.au}

\author[1]{\fnm{Joss} \sur{Bland-Hawthorn}}
\author[1]{\fnm{Thor} \sur{Tepper-Garc\'ia}}
\author[2]{\fnm{Oscar} \sur{Agertz}}
\author[1,3]{\fnm{Elizabeth J.} \sur{Iles}}

\affil*[1]{\orgdiv{Sydney Institute for Astronomy}, \orgname{School of Physics, University of Sydney}, 
\orgaddress{\state{NSW} \postcode{2006}, \country{Australia}}}

\affil[2]{\orgdiv{Lund Observatory, Division of Astrophysics, Department of Physics}, \orgname{Lund University}, \orgaddress{\street{Box 118}, \city{SE-22100 Lund}, \country{Sweden}}}

\affil[3]{\orgdiv{School of Physics}, \orgname{University of New South Wales}, \orgaddress{\state{Sydney, NSW} \postcode{2052}, \country{Australia}}}

\abstract{\added[id= anon3]{The Milky Way is commonly depicted as having four major spiral arms, including the Norma-Outer arm, and a central stellar bar whose properties remain debated.} Masers associated with young high-mass stars have been used to trace a segment of the ``Norma arm" extending from one end of the Galaxy's central stellar bar. Recent work suggests that these masers may reside along a spiral arm or occupy the dominant $x_1$ stellar orbits that give the bar its shape. \added[id =anon3]{We now use a new generation of N-body, hydrodynamic simulations to argue that these million-year old masers instead trace the Galactic bar’s Quadrant I shock front or, equivalently, its dust lane. This implies that the bar dust-lane is star-forming, shedding new light on our understanding of star formation processes in the Milky Way. }
By associating the masers with the simulated shock front, we determine the bar's parameters to be: $a\approx4.3 \pm 0.2$~kpc, $b\approx2.3 \pm 0.2$~kpc, $\phi = 20^{+5}_{-2}\,^\circ$, $\Omega_{p} \lesssim 40 \; \mathrm{km \; s^{-1} \; kpc^{-1}}$, \added[id = anon3]{representing the bar’s semi-major and semi-minor half-lengths, inclination angle and pattern speed, respectively. This raises questions about the Milky Way’s spiral structure and suggests that the Galactic bar is long, slow and star-forming.}}


\maketitle

Maser observations from the BeSSeL Survey and Japanese VERA project have previously been used to trace the ``Norma arm" \added[id = anon]{at its location near the Galactic bar end, beginning at $(x,y) \approx (-2,3)$~kpc and curving anticlockwise to $\approx (-4,  0)$~kpc\,\cite{reid2019trigonometric}, where $x$ and $y$ are Galactocentric coordinates with the Sun at $(-8.5, 0)$.}
In some interpretations, the Norma arm continues around to the far side of the Galactic centre, connecting with the Outer arm $-$  the so-called ``Norma-Outer arm'' \added[id = anon]{(see Figure~1 of\,\cite{reid2019trigonometric} for a schematic of the various spiral arm locations)}. However, recent work suggests that some of the young stars traditionally attributed to Norma actually reside on $x_{1}$ orbits of the Galactic bar, in which case the arm terminates at the bar's end\,\cite{kumar2025expanding}.

Gas structures in the bar region are commonly understood with reference to $x_{1}$ and $x_{2}$ orbits \added[id = anon]{(See \EDref{fig:bar_spiral_schematic})}. Gas flows inward along the  $x_{1}$ orbit family, eventually reaching an innermost `cusped' $x_{1}$ orbit
\cite{binney1991understanding,   morris1996galactic}.
At smaller radii, oval-shaped $x_2$ orbits appear. 
The pericentre of the cusped $x_1$ orbit grazes the apocentre of the outermost $x_2$ orbits, causing shocks to occur\,\cite{binney1991understanding, morris1996galactic, downes1996co, kim2024impacts}, 
\added[id = anon]{which drive gas inward to the $x_2$ family}.

\added[id = anon]{In the Milky Way's bar region, several key structures appear in the gas:} (i) a nuclear disc/ring or Central Molecular Zone (CMZ), which is a dense ring of molecular gas \added[id = anon]{on $x_2$ orbits}\,\cite{morris1996galactic, fux19993d, oka1998co, henshaw2022star}; 
(ii) two distinct dust lanes \added[id = anon]{or connecting arms}, which we argue follow bar-generated shock fronts along the outer edges of the bar; (iii) a gas-poor ``shock cavity'' along the bar, between the dust lanes and the nuclear disc\,\cite{davis2025hi}. These features of the MW's gas distribution are seen in stellar bars of external galaxies: e.g. NGC 1433 observed with the James Webb Space telescope (available at \href{https://science.nasa.gov/asset/webb/ngc-1433-miri-image}{https://science.nasa.gov/asset/webb/ngc-1433-miri-image}).

In the local Universe, star formation (SF) is often suppressed along stellar bars (eg. NGC~3351, 1512, 4303, 5383)\,\cite{downes1996co, kim2024impacts, tubbs1982inhibition,  phillips1996star, reynaud1998kinematics, sheth2000molecular, momose2010star}. 
Specifically, SF is often low along the bar dust lanes, between the nuclear disc and bar ends, despite their substantial reservoirs of dense gas\,\cite{martin1997star, hirota2014wide, renaud2015environmental}. 
This has been explained by several mechanisms, \added[id = anon3]{including bar torques that funnel gas inward, starving the dust lanes of gas needed for star formation\,\cite{newnham2020h, george2019significance, scaloni2024impact, bland2024turbulent}, or bar-driven shocks and shear, which inhibit the growth of gravitational instabilities by driving turbulence\,\cite{kim2024impacts, reynaud1998kinematics, haywood2016milky}.  
}

However, the suppression of SF along local bars is not universal. NGC~7479, NGC~3359, NGC~2903 and NGC~1503, for example, all exhibit significant SF along their dust lanes, likely due to, respectively: a recent minor merger, ample gas supply in a young bar, a broken feathery dust lane structure that weakens shear and shocks, and stars forming in dust spurs but later migrating to dust lanes\,\cite{martin1997star, rozas1999global, zhou2011star,martin1995oxygen,  leon2008hcn, popping2010multiwavelength, block2004gravitational}.

\added[id = anon3]{Furthermore, whether the Galactic bar is long and slow ($R_{\rm b}\,\approx\,5$~kpc and pattern speed $\Omega_{\rm p} \approx 35 - 45 \; \mathrm{km \; s^{-1} \; kpc^{-1}}$)\,\cite{wegg2015structure, monari2019signatures, chiba2021tree}, or short and fast ($R_{\rm b} \approx 3.5$~kpc and $\Omega_{\rm p} \approx 50- 60 \; \mathrm{km \; s^{-1} \; kpc^{-1}}$)\,\cite{hilmi2020fluctuations,vislosky2024gaia, lucey2023dynamically, dehnen2000effect, antoja2014constraints} remains highly debated. Many of the bar properties proposed in the literature are based on observations of Milky Way gas in $(\ell, V_{\rm los})$ $-$ Galactic longitude and line-of-sight velocity $-$ space (eg. see \EDref{fig:dame_image})\,\cite{binney1991understanding, fux19993d, davis2025hi, englmaier1999gas,  weiner1999properties, sormani2015gas3, li2022gas}.  }

\added[id = anon]{$(\ell, V_{\rm los})$ diagrams, formed by the superposition of structures along lines-of-sight, cannot be
deprojected into a unique $(x,y,z, v_x, v_y, v_z)$ distribution. 
For pure-circular motion, the tangent-point circle in $(x,y)$ maps perfectly to the terminal velocity curve in $(\ell, V_{\rm los})$  (see \EDref{fig:TP_circle}). For gas on circular orbits about the Galactic centre, each orbit within the solar circle has two points where a line-of-sight from the Sun is tangent to the orbit (one at each $\ell>0$ and $\ell<0$). The \textit{tangent-point (TP) circle is defined as the locus of these points, forming a circle that connects the Sun and the Galactic centre}. Additionally, the \textit{terminal velocity curve} is the outer envelope of an $(\ell, V_{\rm los})$ diagram, i.e. the maximum $|V_{\rm los}|$ for each longitude (see \EDref{fig:TP_circle}). However, in real galaxies, non-circular motions break this one-to-one correspondence between the TP circle and terminal curve.}

\added[id = anon]{There are numerous flaws with the current interpretation of bar and spiral related gas features in $(\ell, V_{\rm los})$ diagrams. In the following, Galactic Quadrants~I and~IV are defined by $0^{\circ} < \ell < 90^{\circ}$ and $-90^{\circ} < \ell < 0^{\circ}$, respectively, with the Sun at $(-8.5,0)\,\mathrm{kpc}$ (see Fig.~\ref{fig:intro_KE_maps}). In $(\ell, V_{\rm los})$, the Quad~I ``connecting arm'' or dust lane is identified as a ridge (also labelled in \EDref{fig:dame_image}), which begins at the maximum $V_{\rm los}$ for $\ell > 0$, then decreases continuously in $V_{\rm los}$ while increasing continuously in $\ell$, before ending near $\ell \approx 10^{\circ}$, where gas is sparse. This ``connecting arm'' $(\ell, V_{\rm los})$ feature requires a bar that: (i) has a straight Quad~I dust lane, since only a straight lane can increase monotonically in longitude as we move away from the nuclear disc (a curved lane would eventually decrease in $\ell$ as it wraps around the bar; see \EDref{fig:dust_lane_shape}); (ii) hosts a very compact nuclear ring ($-1^{\circ} \lesssim \ell \lesssim 1.5^{\circ}$, $R \approx 0.23$~kpc predicted by tangent-point methods, $R = R_0\sin\ell$)); (iii) is short enough overall to remain confined to $|\ell| \lesssim 10^{\circ}$; and (iv) has a gas-poor bar end, since the connecting arm $(\ell, V_{\rm los})$ ridge terminates at a zone very deficient in gas. This is inconsistent with the standard picture of bars, where bar ends are gas-rich, and often star-forming, with gas being funneled inward along the bar dust lanes. We instead propose that the bar dust lanes lie at larger radii, intersecting the terminal curve near $|\ell| \approx 20^{\circ}$, a region traditionally attributed to spiral arms.}

In this work, we argue that masers, typically linked to regions of massive star formation, align with the Galactic bar’s dust lane in Quad I. \added[id = anon]{We focus on an N-body, hydrodynamical model that is designed to reproduce the Milky Way as much as possible (Model 1, see \nameref{sec:methods} for details)\,\cite{davis2025hi, tep24a}.} \added[id = anon]{Masers result from a population inversion in molecular energy levels, where stimulated emission produces the coherent amplification of radiation at appropriate microwave or radio frequencies\,\cite{reid1981masers, gray1999astrophysical, elitzur1995masers}.} 

\added[id = anon]{From the match between the ``Norma'' masers and the simulated bar shock front in both $(x,y)$ and $(\ell, V_{\rm los})$ space, we:} (i) identify a \textit{curved} dust lane (ii) find that star formation is not fully suppressed in the bar region; (iii) argue that the Norma arm is not a spiral arm, but a shock front along the Galactic bar; (iv) estimate the bar's properties to be $a_{\rm bar} \approx 4.3 \pm 0.2$~kpc,  $b_{\rm bar} \approx 2.3 \pm 0.2$~kpc and $\phi = 20^{+5}_{-2}\,^\circ$.

\added[id = anon]{To begin this paper, the kinematic distance (KD) method is applied to our MW ``surrogate'' simulation, and low KD error zones are used to isolate gas near the bar shock fronts  (see Sec.~\nameref{sec:KE_map} and \nameref{sec:low_KD_shock}). Next, we identify a distinct feature in observed CO diagrams, which we call the ``second peak'' (\EDref{fig:dame_image}), and relate this to the longitude where the TP circle and bar dust lane intersect (Sec.~\nameref{sec:second_peak}). This result is reproduced in a second simulation (Sec.~\nameref{sec:second_sim}). Finally, the simulated shock front is matched with the location of ``Norma'' masers in both $(x,y)$ and $(\ell, V_{\rm los})$ space. This is used to constrain the bar’s size and orientation (Sec.~\nameref{sec:bar_properties}), and to suggest that the Galactic bar's dust lane is star-forming (Sec.~\nameref{sec:SF}).}

\section*{Kinematic Error maps}\label{sec:KE_map}

The kinematic distance (KD) method estimates the distance of a gas cloud from the Sun using its Galactic longitude $(\ell)$ and line-of-sight velocity $(V_{\rm los})$, and by assuming circular motion. \added[id = anon]{Please see \nameref{sec:kin_method} section, in the Methods, for details.}

In Fig. \ref{fig:intro_KE_maps} (bottom row, left panel), we introduce a basic map of kinematic distance error, $(d_{\rm k} - d_{\rm true})/ d_{\rm true}$, for our model at $t = 3.95$~Gyr. Here $d_{\rm k}$ is the kinematic distance estimate, while $d_{\rm true}$ is the actual heliocentric distance taken from the simulation. To put these results in context, the first row of Fig. \ref{fig:intro_KE_maps} displays a stellar density projection (left) and gas density projection (right) for our simulation at the same time. The model features a stellar bar, which we have oriented at a $25^{\circ}$ angle relative to the $x = 0$ axis. In the bottom, right corner of Fig \ref{fig:intro_KE_maps}, we isolate zones of low KD error ($|(d_{\rm k} - d_{\rm true})/d_{\rm true}| < 0.02$) in green.

\section*{Low KD error zone along bar shock front}\label{sec:low_KD_shock}

In the gas density projection of Fig. \ref{fig:intro_KE_maps} (top right panel), the \textit{bar shock fronts}, or dust lanes, are labelled. These dense structures occur along the leading sides of the bar, where inflowing gas is compressed by bar-driven shocks.

\added[id = anon]{
In Fig.~\ref{fig:bar_shock_front}, we zoom in on low KD error zones in the bar region (left panel), and map these points into $(\ell, V_{\rm los})$ space (right panel). The low KD error zone adjacent to the bar's leading edge and nuclear spiral in Quad I (presented with the \texttt{magma}, or dark purple to pale yellow, colormap) appears as a loop-like structure at $\ell > 0$ in $(\ell, V_{\rm los})$. The \texttt{magma}-coloured points in $(\ell, V_{\rm los})$ exhibit a clear ``peak'' -- this is labelled as the ``second peak'' in Fig.~\ref{fig:bar_shock_front} and it maps to an intersection point between the tangent-point circle and the bar dust lane (see pink points)}.

\section*{The Second Peak Feature}\label{sec:second_peak}

\added[id = anon]{\EDref{fig:dame_image} presents the $(\ell, V_{\rm los})$ distribution of observed carbon-monoxide (CO) in the MW. We label a \textit{first} and \textit{second} ``peak'' feature in Quad I $(0^{\circ} < \ell < 90^{\circ})$. These are defined as a sharp change in $|V_{\rm los}|$ along the terminal velocity curve, such that we can identify an obvious visual peak. In Quad IV $(-90^{\circ} < \ell < 0^{\circ})$, three analogous ``peak-like'' features are also present, but are not the focus of this work.} \added[id = anon2]{A more detailed definition of the ``peak'' features is given in the Methods, 
Sec.~\nameref{method:sim_peak}.}

\added[id = anon]{In Figure~\ref{fig:magenta_grid}, we compare $V_{\rm los}$ along the tangent-point (TP) circle (magenta curves) against the outer envelope of simulated $(\ell, V_{\rm los})$ diagrams (grey distributions). This analysis is performed for all barred snapshots in our MW ``surrogate'' model, with results for representative times between $t = 1.44$–$3.95$~Gyr shown in Fig.~\ref{fig:magenta_grid}. Please see Methods Sec.~\nameref{method:TP} for details.}

\added[id = anon]{The grey simulated $(\ell, V_{\rm los})$ distributions \textit{all} have at least two clear ``peak-like features'' in each Quad I and Quad IV, marked with the dark blue scatter points in Fig. \ref{fig:magenta_grid} (we call these the ``simulated true peaks''). Additionally, the magenta curves consistently display two equivalent ``peaks'' in each quadrant (``TP peaks''). The simulated true peaks and TP peaks always lie very close to each other.}

\added[id = anon]{The magenta points are taken directly from the TP circle of a $V_{\rm los}$ projection in $(x,y)$ space. The top left panel of \EDref{fig:hunter_image} gives an example. From these $V_{\rm los}$ projections, we know that the ``TP second peak'' occurs when the TP circle intersects the bar dust lane. Hence, we consider that the true simulated peaks may also result from the bar. }

\added[id = anon]{Low KD error zones were previously used to approximately trace the gas beside the bar shock front (Sec.~\nameref{sec:low_KD_shock}). In panels (a) and (b) of Fig. \ref{fig:periodograms}, we compute KD errors and $V_{\rm los}$ values along the TP circle, and track their evolution over time in the simulation. In both panels (a) and (b), two near-horizontal ridges stand out at low longitudes ($|\ell| \approx 10^{\circ}$), marked by the white-dashed lines, and characterised by moderate KD errors ($0.15 \lesssim |(d_k - d_{\rm true})/d_{\rm true}| \lesssim 0.2$) and moderate $V_{\rm los}$ ($\approx 120$~km~s$^{-1}$). On either side of these \textit{white-dashed ridges} lie zones of both extremely low KD error (appearing dark blue to black) and very high $V_{\rm los}$ (dark red in Quad I, or dark blue in Quad IV). We label these zones of high $V_{\rm los}$ in panel (b) as the ``TP first peak'' and ``TP second peak'', reflecting their location at the first and second peaks in the previously discussed magenta curves (Fig.~\ref{fig:magenta_grid}).} 

\added[id = anon]{The \textit{white-dashed ridges} and the low KD error/high $V_{\rm los}$ zones on either side of them arise from a consistent pattern in our simulation: moving from $\ell \approx 15^{\circ}$ toward $0^{\circ}$, the KD error transitions from an extremely low-error region (tracing the bar shock front) to the moderate-error ridge (just beyond the shock), before decreasing back down to very low error values near $\ell \approx 0^{\circ}$ (centre of the nuclear ring). Furthermore, the two white-dashed ridges move away from each other with time, consistent with our simulated bar growing over time. Importantly, throughout the simulation's duration,  along the TP circle,  zones of high $V_{\rm los}$ (ie. ``peak features'') on either side of these white-dashed ridges consistently coincide with low KD error zones, which we have shown to trace the bar.}

\added[id = anon]{In panel (c) of Fig. \ref{fig:periodograms} (bottom left), we isolate the low KD error $|(d_k - d_{\rm true})/d_{\rm true}| < 0.02)$ zones from panel (a) and present the result with the green scatter points. Next, with the black curve, we trace along the low KD error zone that should approximately follow the bar shock front. Hence, the black points should give a reasonable estimate of the longitude value for the TP intersection point with the bar shock front. This was also verified by overlaying the longitude lines for the black curve on $(x,y)$ gas distributions, and visualising the result.}

\added[id = anon]{In panel (d) of Fig.~\ref{fig:periodograms} (bottom right), we show the longitude values of the \textit{true} first, second, and occasionally third peaks in the simulated $(\ell, V_{\rm los})$ diagrams (see Methods Sec.\ref{method:sim_peak}). Examples of these peak locations are also shown as the dark blue points in Fig.~\ref{fig:magenta_grid}.} \added[id = anon]{Additionally, we overlay the black curve that was previously used to locate the low KD error zone adjacent to the bar shock front. In panel (d), we see that this black curve goes almost perfectly through the blue scatter points that identify the Quad I ``second peaks''. \textit{This indicates a strong relationship between the location of the bar shock front and the longitude value of the ``second peak'' in $(\ell, V_{\rm los})$ diagrams. This relationship appears as a pattern across all barred snapshots in the simulation.}}

\subsection*{Bar-spiral attachment}

 \added[id = anon]{\EDref{fig:second_peak_fluctuations} shows a striking resemblance between the amplitude of the $m=2$ fourier mode (top row) and the $V_{\rm los}$ coordinate of the Quad I second peak (middle row). Oscillations in $A_2$ have long been attributed to spiral arms attaching and detaching from the bar\,\cite{debattista2000constraints, pfenniger2023five}. This suggests second peak $V_{\rm los}$ also fluctuates as spiral arms attach/detach to the bar. Furthermore, spiral arm strength (density contrast and shocks) decreases as the model evolves, which is reflected in the dampening of the $A_2$ and second peak $V_{\rm los}$ oscillations over time. While attached spiral arms can cause fluctuations in second peak $V_{\rm los}$, baseline $V_{\rm los}$ values (about which the fluctuations occur) appear to be set by the bar. In addition, second peak $V_{\rm los}$ depends on the bar pattern speed, decreasing as $\Omega_p$ decreases (see Fig.~11 of\,\cite{davis2025hi}). Overall, the ``second peak'' longitude coordinate is most valuable for constraining bar size and angle; while ``second peak'' $V_{\rm los}$ value tells us more about bar pattern speed and spiral-arm attachment.}

\section*{Comparison to a second simulation }\label{sec:second_sim}

\added[id = anon]{In \EDref{fig:Liz_sim}, we present the $(x,y)$ and $(\ell, V_{\rm los})$ gas density distributions at $t = 920$~Myr in a second barred galaxy simulation (Model 2, see Sec.~\nameref{sec:intro_liz_model} of the Methods for details). We mark the ``second peak'' locations in the $(\ell, V_{\rm los})$ diagram with dark blue dashed lines, and present the corresponding lines of longitude in $(x,y)$ space. In $(x,y)$, these lines clearly intersect the pink tangent-point circle near the bar dust lanes. This demonstrates that the link between the $(\ell, V_{\rm los})$ ``second peaks'' and the bar dust lanes is also present at $t = 920$~Myr in this barred galaxy simulation.}

\section*{The Norma arm traces Galactic bar shock front}\label{sec:norma}

We take the CO $(\ell, V_{\rm los})$ observations (\EDref{fig:dame_image}), visually identify the Quad I ``second peak'', and estimate its approximate location in $(x,y)$ space, using the TP-dust lane intersection method and kinematic distance methods (see Methods Section~\nameref{sec:test}). In Fig.~\ref{fig:observation_comparison}, \textit{the gold scatter point represents both the coordinates of this CO Quad I ``second peak'' in $(\ell, V_{\rm los})$ space (middle row) and its estimated location in $(x,y)$ (top row)}.  
Similarly, $(x,y)$ predictions for the Quad IV ``second'' and ``third'' peaks are shown with the black scatter points. Results are summarised in Extended Data Table \ref{tab:CO_peak_estimates}.

\added[id = anon]{The top left panel of Fig. \ref{fig:observation_comparison} presents the high-mass star-forming regions with measured trigonometric parallaxes from\,\cite{reid2019trigonometric}. The red points in both $(x,y)$ and $(\ell, V_{\rm los})$ represent masers that have been identified with the ``Norma'' spiral arm\,\cite{reid2019trigonometric}}.

\added[id = anon]{The top right panel of Fig.~\ref{fig:observation_comparison} shows the simulated $(x,y)$ gas density projection at $t = 3.95$~Gyr. The data points from observations -- specifically, the trace of the red masers, combined with the gold point -- have a very similar curvature to our simulated bar shock front and could easily trace a Galactic bar shock front that is slightly larger than the one in our model here.}

\added[id = anon]{The second row of Fig.~\ref{fig:observation_comparison} shows the corresponding $(\ell, V_{\rm los})$ distributions. The left-middle panel presents the observed HI emission from\,\cite{reid2019trigonometric}. The gold point, representing the CO ``second peak'', is also located at a clear peak in HI. See Fig.~15 of\,\cite{fux19993d} for a simple comparison of CO and HI in the Milky Way\,\cite{fux19993d}.}

\added[id = anon]{In the middle-right panel of Fig.~\ref{fig:observation_comparison}, we present the simulated $(\ell, V_{\rm los})$ distribution for $t = 3.95$~Gyr. Once again, the path of the red masers and gold scatter point is consistent with a Galactic bar that is slightly larger in size and slower in pattern speed, than the bar present in our simulation at $ t= 3.95$~Gyr. The gold point (second peak from observations) has a slightly larger $\ell$ and lower $V_{\rm los}$ than our simulated ``second peak'' does consistent with a larger Galactic bar size and slow bar pattern speed $\Omega_p \lesssim 40 \; \mathrm{km s^{-1} kpc^{-1}}$, respectively. Changes in spiral attachment could also decrease the ``second peak'' $V_{\rm los}$ to make it consistent with observations.}

\added[id = anon]{Finally, the bottom left panel is the hypothesised stellar density distribution from\,\cite{wegg2015structure} based on red clump giants, while the bottom right panel shows the simulated stellar density distribution at $t = 3.95$~Gyr. The gold point (our predicted TP--bar dust-lane intersection), initially suggests a slightly wider bar than in\,\cite{wegg2015structure}. However, star-forming regions can be offset to the leading side of the stellar bar by up to $15^\circ$\,\cite{martin1997star, sheth2002molecular}, so this offset remains consistent with the stellar bar of\,\cite{wegg2015structure}.}

\section*{Constraining Galactic bar size and angle}\label{sec:bar_properties}

\added[id = anon]{If the Norma arm traces the bar-driven shock in Quadrant~I, this tightly constrains the bar length. In Fig.~\ref{fig:MW_bar_length} we scale and rotate the \(t=3.95\) gas density map (Fig.~\ref{fig:observation_comparison}) to align the simulated shock front with the Norma masers and the approximate TP--bar intersection (gold point), highlighting the associated low KD-error zone in orange. By visual inspection, we draw cyan dashed lines across the semi-major and semi-minor axes of the bar, or more precisely across the estimated location of the cusped $x_1$ orbit.}

\added[id = anon]{The best alignment occurs for $a\approx4.3 \pm 0.2$~kpc, $b\approx2.3 \pm 0.2$~kpc and $\phi = 20^{+5}_{-2}\,^\circ$ (see Sec. \ref{method:bar_length} for details on bar length measurement and uncertainties). This is consistent with $R_{\rm b}=4.6\pm0.3$~kpc\,\cite{wegg2015structure}, supporting the argument for a long bar.}

\section*{Star formation in the bar}\label{sec:SF}

The Norma masers (predominantly H$_2$O and CH$_3$OH\,\cite{reid2019trigonometric}) trace very young ($\ll 1$~Myr) star-forming regions that remain close to the gas clouds in which they formed because they have not had sufficient time to migrate away\,\cite{reid2019trigonometric, kumar2025expanding}. \added[id = anon]{Hence, the association of ``Norma'' masers with the Galactic bar shock front suggests that SF is occurring along the Galactic bar dust lane in Quad I.}

\added[id = anon]{Recent work shows that ``strong'' bars with $\mathscr{R}=R_{\rm CR}/a_{\rm bar}>1.4$\,---\,where $R_{\rm CR}$ is the corotation radius\,---\,exhibit the highest SF in the bar region, with SF also able to occur along the bar dust lanes\,\cite{mengistu2026effects}. In this context, ``strong'' refers to Galaxy Zoo morphological classifications, largely reflecting bar size relative to the disc (different to e.g. bar torque, $A_2$). Furthermore, $\mathscr{R} > 1.4$ is often used to classify bars as ``slow', although we acknowledge that $\mathscr{R}$ is not always a good diagnostic of pattern speed $\Omega_p$\,\cite{davis2025hi} (see their Fig.~21).
Our hypothesised Galactic bar ($a\approx4.3$~kpc, $\Omega_p\lesssim40~\mathrm{km~s^{-1}~kpc^{-1}}$) is consistent with such extended, ``slow'' bars, supporting 
the picture of SF in a bar that may also contribute to quenching its host galaxy.}

\clearpage

\begin{figure}
    \centering
    \includegraphics[width=\textwidth]
    {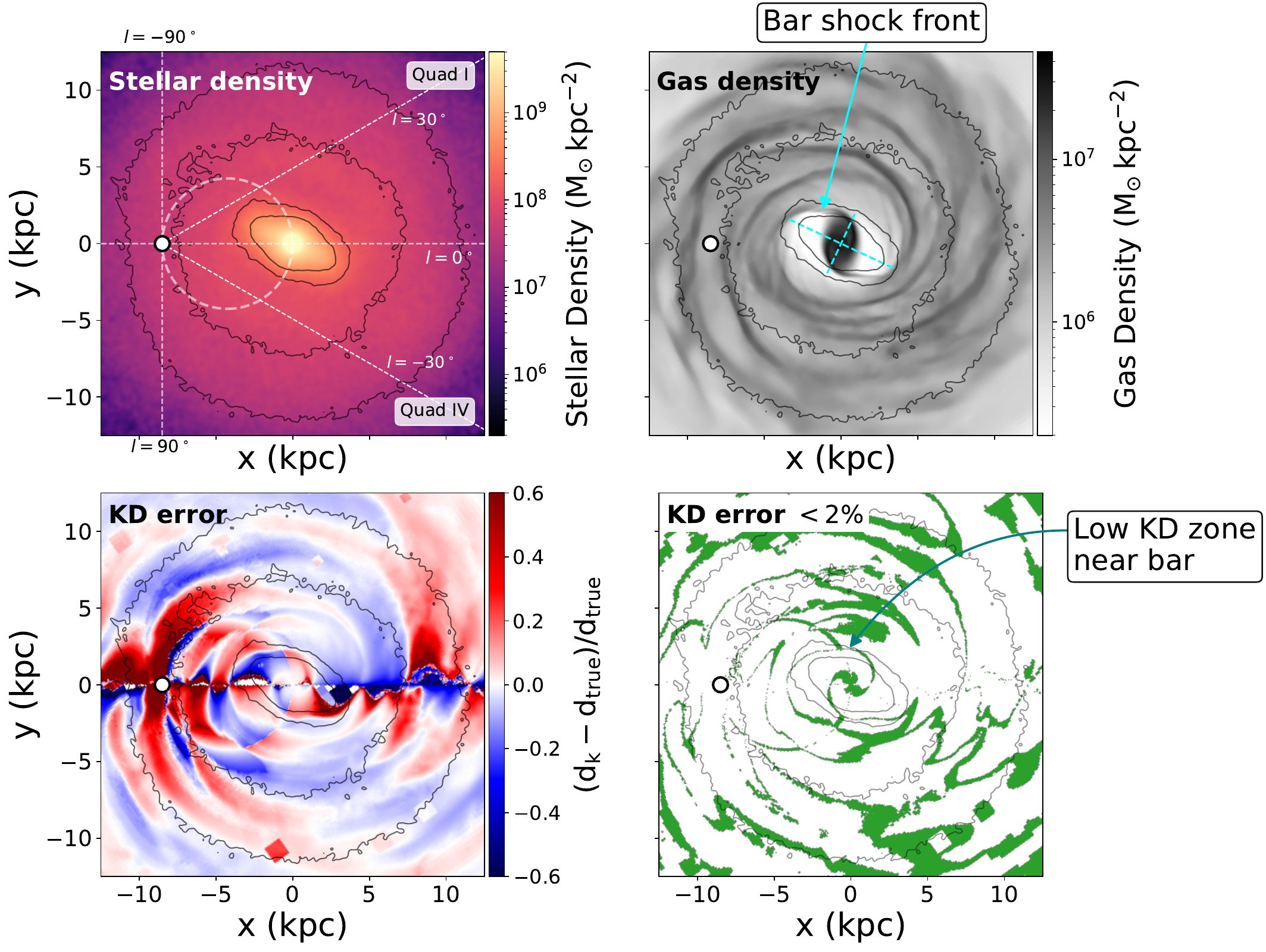}
    \caption{\textbf{Kinematic error maps in our Milky Way-like simulation.} Top-left shows the stellar density distribution at $t = 3.95$~Gyr in the model. The bar has been oriented at $25^{\circ}$ relative to the x-axis. Galactic Quadrants I ($0^{\circ} < \ell < 90^{\circ}$) and IV ($-90^{\circ} < \ell < 0^{\circ}$) are labelled. The dashed white lines represent several key lines-of-longitude $\ell = 0^{\circ}, \pm 30^{\circ}, \pm 90^{\circ}, 180^{\circ}$. Furthermore, the white dashed circle marks the location of the tangent-point circle. The top-right panel shows the corresponding gas density distribution. The cyan dashed lines mark the approximate imprint of the stellar bar on the gas distribution. Bottom-left panel shows the kinematic distance error map. The \added[id = anon]{bottom-right} panel displays low KD error zones  ($|(d_{\rm k} - d_{\rm true})/ d_{\rm true}| < 0.02$) in green. The white scatter point with black outline at $(x,y) = (-8.5, 0)$~kpc represents the Sun. The black contours overlaid on all panels are stellar density contours, taken from the top-left stellar density distribution.}
    \label{fig:intro_KE_maps}
\end{figure}

\begin{figure}
    \centering
    \includegraphics[width=\textwidth]
    {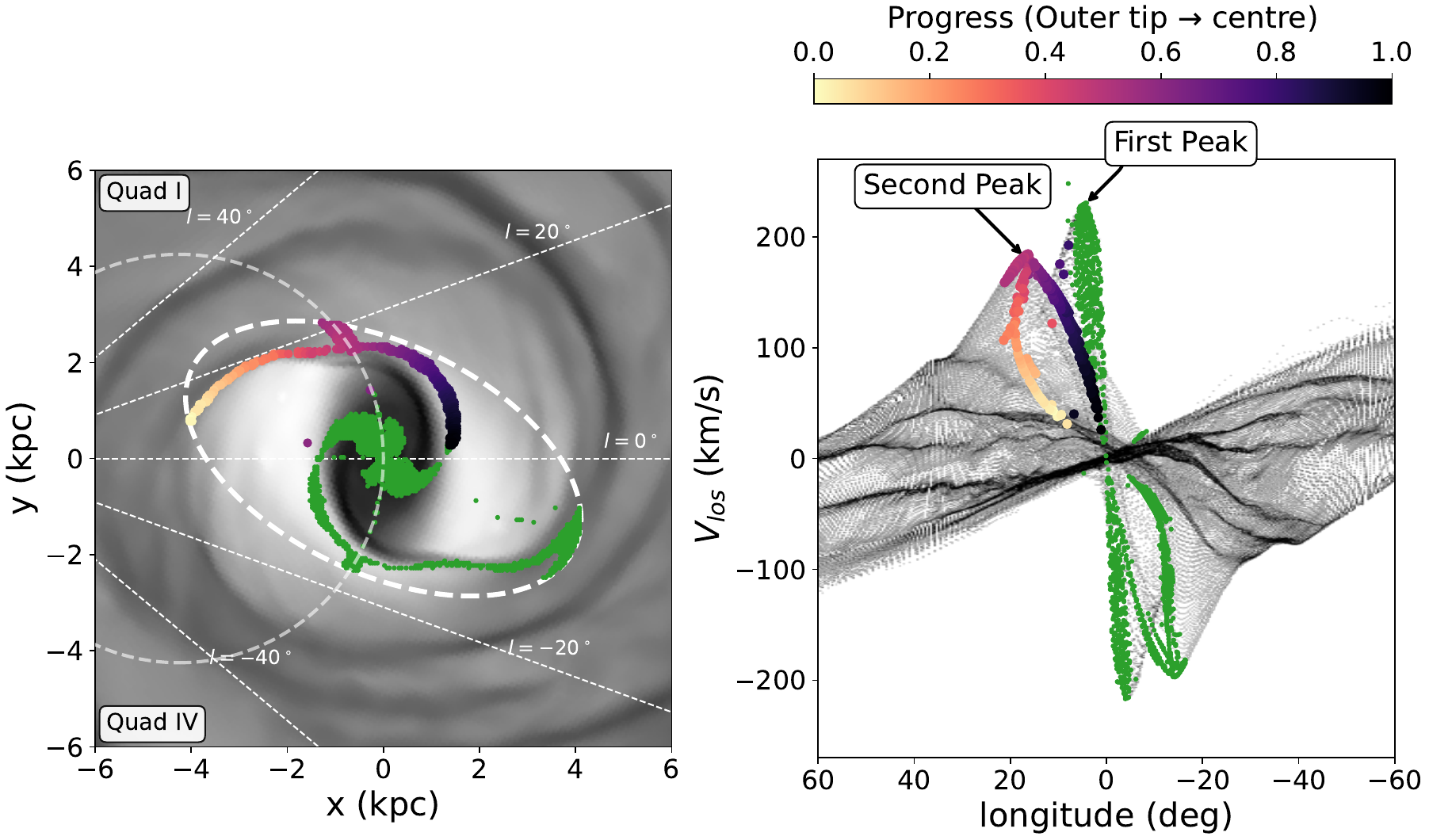}
    \caption{\textbf{Tracing the bar shock front in simulated gas maps.} $(x,y)$ (left panel) and $(\ell, V_{\rm los})$ (right panel) gas density distributions at $t = 3.95$~Gyr in the model. We isolate low KD error zones approximately within the bar region using the white ellipse (drawn by eye), and for the most part, these points are presented in the green. However, for the low KD error zone along the bar's leading edge and nuclear spiral in Quad I -- a focus of our work -- we colour these with the \texttt{magma}, or dark purple to pale yellow, colormap. As shown with the colorbar, this structure is coloured pale yellow near the bar end and then progresses to dark purple as we move toward the bar centre. Hence, we can more easily visualise the corresponding location of individual points in $(\ell, V_{\rm los})$ space. The ``First peak'' and ``Second peak'' features in Quad I of this simulated $(\ell, V_{\rm los})$ distribution are labelled. In $(x,y)$, the tangent-point circle and some key lines-of-longitude are given by the white dashed circle and white dashed lines, respectively. We emphasise that the intersection of the TP circle with the bar dust lane occurs at the location of the ``second peak'' feature in $(\ell, V_{\rm los})$ (see pink points).}
    \label{fig:bar_shock_front}
\end{figure}

\begin{figure}
    \centering
    \includegraphics[width=\textwidth]{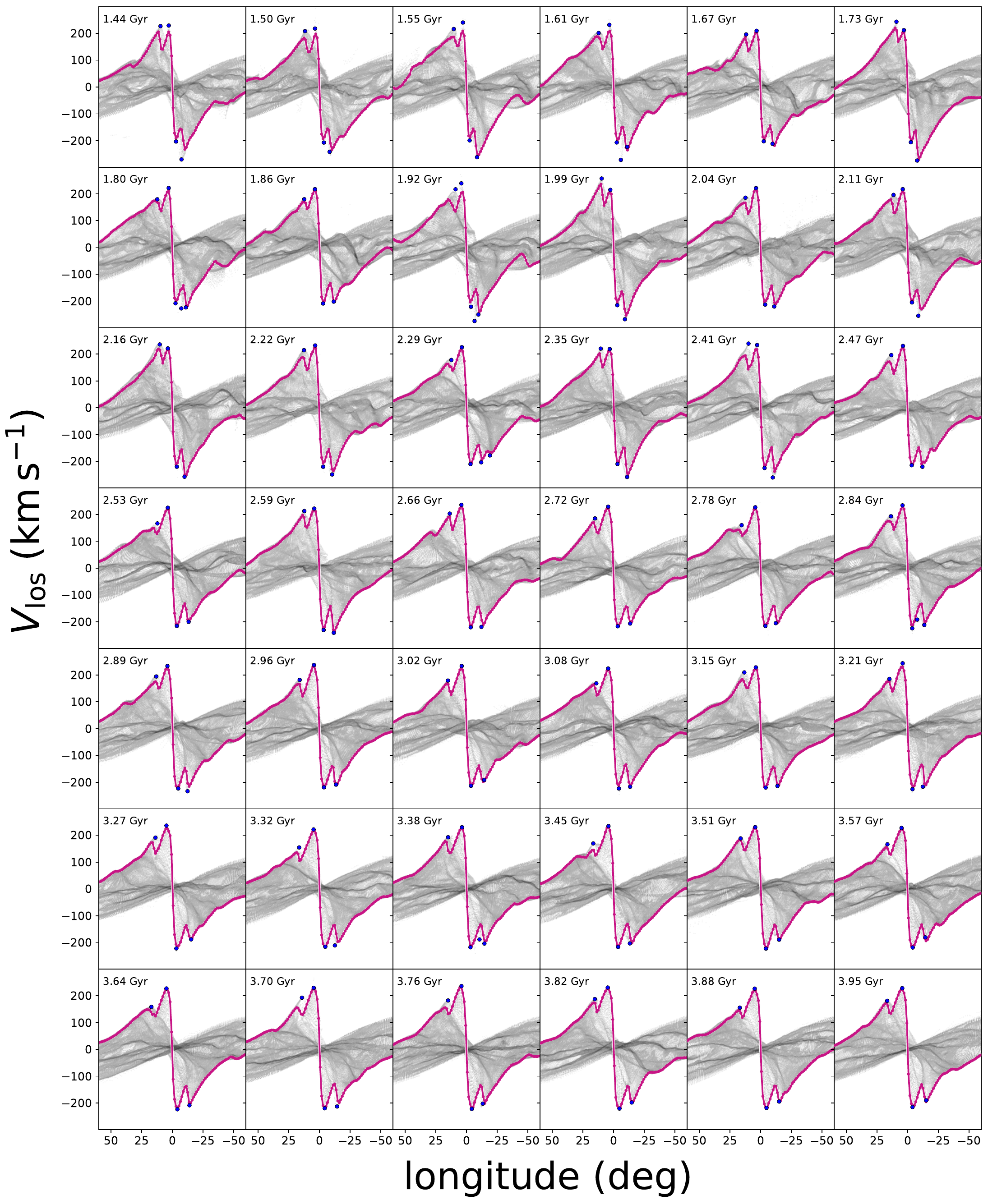}
    \caption{\added[id = anon]{\textbf{Comparing terminal velocities with tangent-point circle velocities.} $(\ell, V_{\rm los})$ gas distributions (shown in grey) for our Milky Way ``surrogate'' simulation for selected snapshots between $t = 1.44-3.95$~Gyr. Blue points mark the true first and second ``peak'' features for these simulated $(\ell, V_{\rm los})$ distributions. Occasionally, there is a third peak feature marked as well. The magenta curves show the mean $V_{\rm los}$ and $\ell$ values computed around the TP circle (see Methods Section \ref{method:TP} for details). Overall, the magenta curves have a very similar shape to the outer envelope of the corresponding grey $(\ell, V_{\rm los})$ distributions.
    }}
    \label{fig:magenta_grid}
\end{figure}

\begin{figure}
    \hspace*{-2.5cm}
    \includegraphics[width=1.4\textwidth]{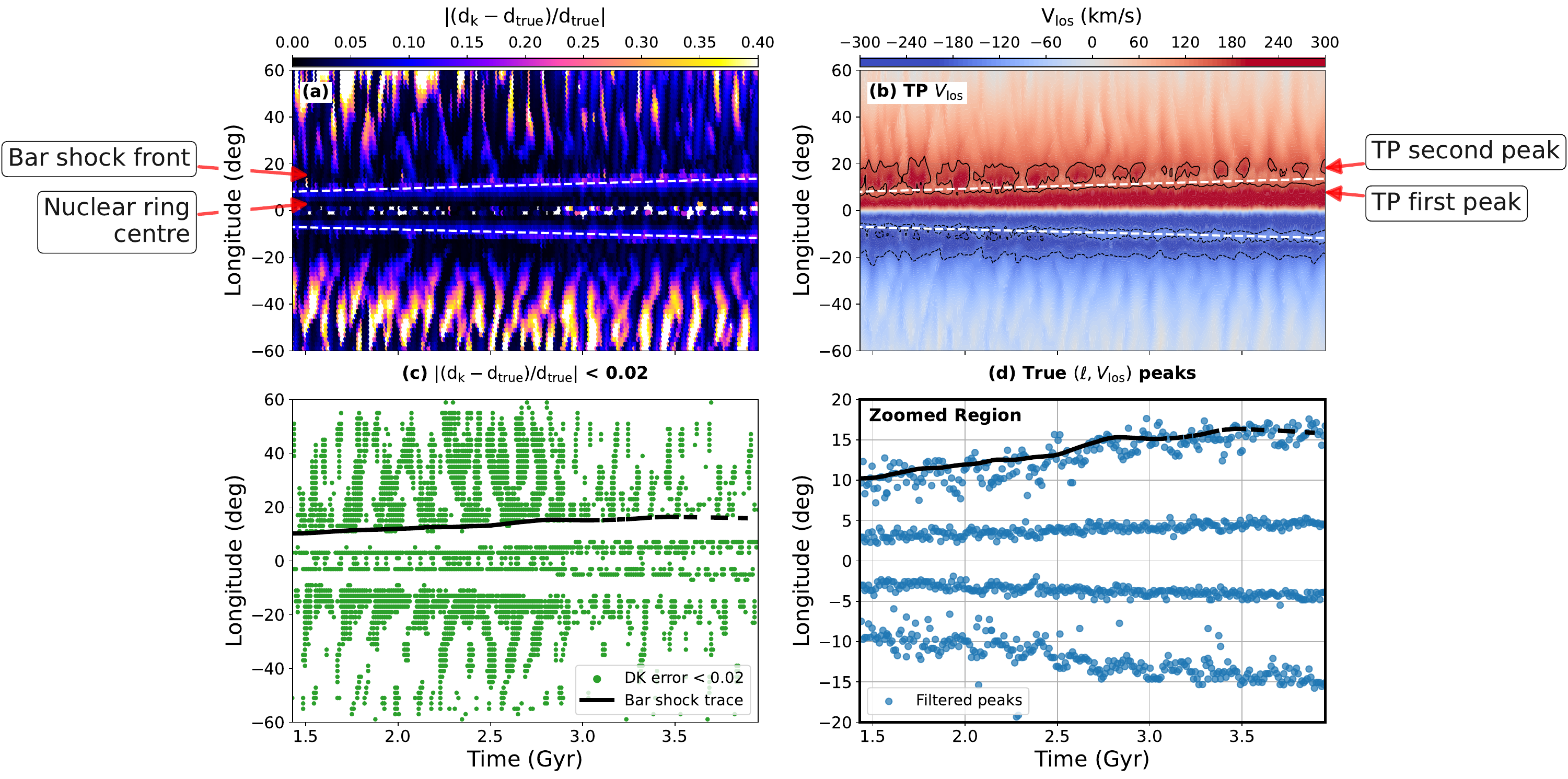}
    \caption{\added[id = anon]{\textbf{Second peak longitudes and the bar shock front location.}
    Panel (a, top left): We compute KD errors around the TP circle, and track their evolution over time in the simulation. We omit early times in the simulation because the bar has not yet formed. In panel a, there are two distinct ``moderate error ridges'' at low longitudes, which we have marked with the white dashed lines. Focusing on Quad I, there is a clear low error zone (mostly black) immediately above the white-dashed ridge, and also immediately below it. These low error zones occur adjacent to the bar dust lane, and at the centre of the nuclear ring, respectively (as labelled). Panel (b, top right): We compute $V_{\rm los}$ around the TP circle, and track its evolution over time. Two ridges of moderate $V_{\rm los}$ appear, which are marked by the same white dashed lines as in panel a. On either side of these white dashed ridges, $V_{\rm los}$ is at its highest values. These high $V_{\rm los}$ zones correspond to the first and second peak features in the \textit{tangent-circle} $V_{\rm los}$ profiles (the mean $V_{\rm los}$ computed along the TP circle, as a function of $\ell$). Panel (c, bottom left): We isolate low KD error zones in \added[id = anon2]{green}. \added[id = anon]{We draw a black curve} through the low KD error zone adjacent to the bar shock front in Quad I. Panel (d, botom right): Blue points show the \textit{true} first and second peak $\ell$ values, which we have computed using the simulated $(\ell, V_{\rm los})$ gas distributions. We overlay the black curve from panel c. The black curve, originally found using the bar shock front location, shows strong agreement with the longitudes of the Quad I ``second peak'' here.}}
    \label{fig:periodograms}
\end{figure}

\begin{figure*}
    \centering
    \vspace{-36pt}   
    \includegraphics[width=\textwidth]{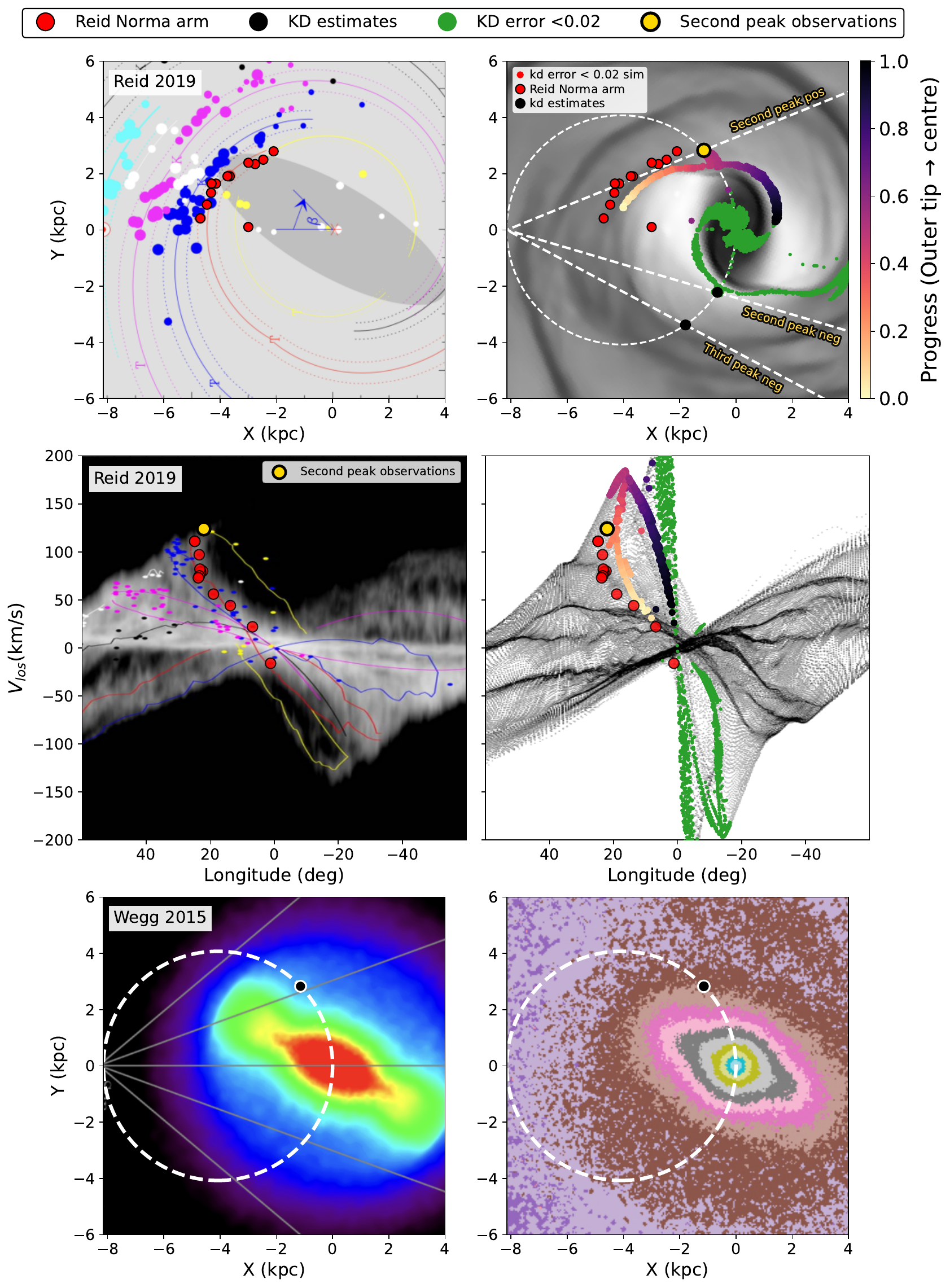}
    \caption{\textbf{Matching the simulated bar shock front with Norma masers.} The top-left panel displays the distribution of HMSFRs, adapted from\,\cite{reid2019trigonometric},
    $\copyright$ AAS. Reproduced with permission. The maser locations are identified with four major spiral arms, coloured red (Norma-Outer), blue (Scutum-Centaurus), magenta (Sagittarius-Carina) and cyan (Local). We have highlighted the red scatter points associated with the Norma arm. The grey ellipse is used by\,\cite{reid2019trigonometric} to indicate the approximate location of the long bar, but is not particularly important for this work. White points mark sources that could not easily be identified with any of the major spiral arms. The Galactic Centre location is given by the red star symbol at $(0,0)$ and the red bull's eye at $(-8.15,0)$~kpc is the Sun. $B$ defines the Galactic azimuth, which is the angle between the Sun-Galactic Centre line, and the line from the Galactic Centre to the source of interest. The top-right panel shows the simulated $(x,y)$ gas density distribution at $ t = 3.95$~Gyr. The white-dashed circle marks the tangent-point circle, and the gold scatter point represents our prediction for the TP circle--bar dust lane intersection point, made using the Quad I ``second peak'' from the CO $(\ell, V_{\rm los})$ observations of\,\cite{dame2001milky}. The black scatter points represent the corresponding predictions made with the Quad IV ``second'' and ``third'' peaks. The white-dashed lines mark the lines-of-longitude corresponding to these Quad I and Quad IV peaks from the CO $(\ell, V_{\rm los})$ observations. Green and \texttt{magma}-coloured (dark purple to pale yellow) scatter points represent the low KD error $( < 2\%)$ zones in the bar region. Red points show the Norma masers from\,\cite{reid2019trigonometric}. In the second row, we have the observed HI $(\ell,  V_{\rm los})$ distribution, also adapted from\,\cite{reid2019trigonometric},
    $\copyright$ AAS (Reproduced with permission). The ``second peak'' from the CO observations\,\cite{dame2001milky} is marked with the gold point. The red points show the Norma masers\,\cite{reid2019trigonometric}. The middle-right panel shows the simulated $(\ell, V_{\rm los})$ at $ t = 3.95$~Gyr. In the bottom panel, we have the estimated stellar density distribution from\,\cite{wegg2015structure}, and the $(x,y)$ stellar density distribution from our model.Once again, the white-dashed circle marks the TP circle, and the grey lines are several key lines-of-longitude, namely $\ell = 0^{\circ}, \pm 20^{\circ}, \pm 40^{\circ}$.}
    \label{fig:observation_comparison}
\end{figure*}

\begin{figure}
    \centering
    \hspace{-1.4cm}
    \includegraphics[width=1.1\textwidth]{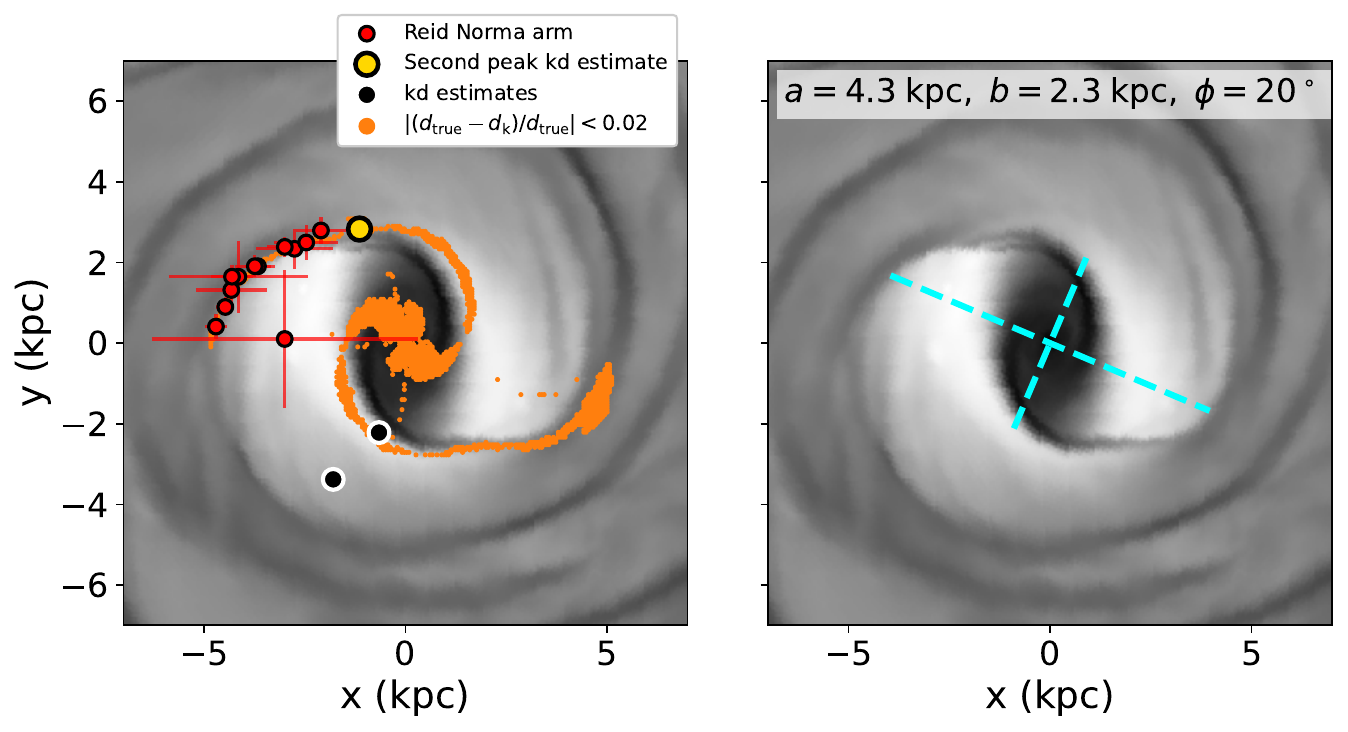}
    \caption{\textbf{Estimating the Galactic bar properties using Norma masers.} We have rotated and scaled the $(x,y)$ gas density distribution from the $t = 3.95$~Gyr snapshot of the simulation, such that the simulated shock front lines up with the observations as closely as possible. The orange points represent the rotated and scaled low KD error zone from the simulation. The red scatter points are the Norma masers, \added[id = anon]{with uncertainties found using the parallax uncertainties of\,\cite{reid2019trigonometric}.} \added[id = anon]{The gold scatter point in Quad I, and} black scatter points in Quad IV are our predictions for intersection points between the TP circle and the bar dust lanes, or alternatively intersection of the TP circle with a spiral arm. On the right panel, we overlay cyan dashed lines across the approximate location of the semi-major and semi-minor axis for the low density shock cavity in the scaled simulation snapshot. In this way, we estimate the Galactic bar size. The text at the top of the right panel, $a = 4.3$~kpc, $b = 2.3$~kpc, $\phi = 20^{\circ}$, are the values we estimate for the Galactic bar parameters when scaling and rotating the simulation snapshot to best match the masers. $a$ and $b$ are estimated using the half-length of the cyan dashed lines in this figure, with uncertainties later taken into account.} 
    \label{fig:MW_bar_length}
\end{figure}

\backmatter
\clearpage
\section*{Methods}\label{sec:methods}

\subsection*{A Milky Way surrogate model}\label{sec:models}

\added[id = anon3]{Model 1 is an N-body, hydrodynamic ``surrogate'' model of the Milky Way (MW), which is designed to reproduce as many observed properties of the Galaxy as possible. Most of the analysis in this paper is performed on this simulation. Detailed descriptions of the model and comparisons with Milky Way observations are presented in \cite{tep24a} and \cite{davis2025hi}. In summary, this ``surrogate'' model consists of four key components: a stellar disc, a central `classical' stellar bulge, a dark-matter halo and a gaseous disc. The properties of each component are summarised in Supplementary Table \ref{tab:s3b}. All components experience self-gravity and react to any changes in the gravitational potential. The gas is treated as monoatomic and ideal, and includes star formation, radiative heating and cooling, stellar feedback, and chemical enrichment.}

\added[id = anon3]{This simulation is built with the {\sc Nexus} framework\,\cite{tep24a}; the initial conditions are generated using an extension of the {\sc agama} library\,\cite{vas19a}, which incorporates both collisionless components and gaseous components. The model was evolved with {\sc Ramses}\,\cite{tey02a}, which is a grid-based, N-body/hydrodynamical code with adaptive mesh refinement (AMR). The simulation's total duration is $\approx 4.3$~Gyr and it is confined to a cubic box with side length 600 kpc. The maximum refinement level is $\ell = 14$ and the maximum spatial resolution is \mbox{$~74$~pc.}}

\subsection*{An alternative barred-galaxy model }\label{sec:intro_liz_model}
\added[id = anon]{We repeat our analysis on a second barred galaxy simulation. We call this Model 2. This is an $N-$body SPH (smoothed particle hydrodynamics) isolated disk galaxy simulation (IsoB) from\,\cite{iles2022}.  The model is fine-tuned to reproduce the nearby galaxy \textsc{ngc4303} as much as possible (using its rotation curve, stellar and gas mass fractions)\,\cite{iles2022}. The simulation consists of particle components for the stellar disk, stellar bulge, dark matter halo and gas disk. It features star-formation and stellar feedback, including: UV and photoelectric heating, metal cooling\,\cite{shen2010} and super-bubble\,\cite{keller2014superbubble} feedback\,\cite{iles2022}. The initial gas particle mass resolution is $1044$\,M$_\odot$ and gravitational softening length for the gas is 10\,pc. This disc was originally produced to compare the star forming properties of bars in simulations with observations, but it has since been applied to studies of radial migration\,\cite{iles2024} and kinematics\,\cite{joshi2026}. It is evolved for a short period of time $\approx 1$~Gyr, but a bar does form and begin to evolve. We focus on $t = 920$~Myr in the model. 
\added[id = anon2]{To select this snapshot, we visually inspected movies of $(x,y)$ stellar and gas density distributions, evolving across the entire simulation's duration. The model begins with a smooth, axisymmetric disc. By $t \approx 500$~Myr, a bar has clearly formed in the stars. It is not until later in the model, once the stellar bar has evolved for some time, that the bar shock fronts or dust lanes appear well-defined in the corresponding gas distribution. At $t = 920$~Myr, a stellar bar is present, and the bar shock fronts appear clearly in the gas distribution. Hence, we applied our method to this snapshot. We were immediately successful in using the second ``peaks'' to locate the bar dust lanes.} Previous studies have assessed the bar length in this snapshot to be 2.73\,kpc (by eye)\,\cite{iles2025} or from 1.47-2.75\,kpc (by automated methods)\,\cite{iles2026}.}
\added[id = anon]{All analysis for this model is also performed using a low-latitude ($|b| < 2^{\circ}$) cut.}


\subsection*{The Kinematic Distance Method}\label{sec:kin_method}

Accurately determining distances to the gaseous component of the Milky Way remains a significant challenge. At heliocentric distances less than approximately 10 kpc, trigonometric parallaxes to maser sources are widely regarded as the ``gold standard'' distance method, providing the most accurate measurements\,\cite{reid2022accuracy, wenger2018kinematic, reid2014trigonometric}. Beyond this range, and on the far side of the Galactic centre especially, kinematic distance estimates are more commonly used\,\cite{anderson2012green, hunter2024testing}.

The KD method estimates the galactocentric radius for a parcel of gas by combining its line-of-sight velocity, $V_{\rm los}$, with an assumed Galactic rotation curve, $(R, \Theta(R))$\,\cite{hunter2024testing, van1954spiral, oort1958galactic}:

\begin{equation}\label{equ:kin_R}
    R = R_{0} \sin(\ell) \frac{\Theta(R)}{\Theta_{0}\sin(\ell) + V_{\rm los} }
\end{equation}

where $R_{0}$ is the sun's distance from the Galactic centre, $\Theta_{0}$ is the sun's circular velocity, $\ell$ is Galactic longitude, and $V_{\rm los}$ is line-of-sight velocity. The estimated galactocentric radius is then used to compute a kinematic distance: 
\begin{equation}\label{equ:kin_distance}
    d_{\rm k} = R_{0}\cos (\ell) \pm \sqrt{R^{2} - (R_{0} \sin(\ell))^{2}}
\end{equation}

When $V_{\rm los} < -\Theta_{0}\sin\ell$, the estimated Galactocentric radius becomes negative, and these points are therefore excluded from our analysis. Furthermore, $d_{k}$ is undefined when $[R^{2} - (R_{0} \sin(\ell))^{2}] < 0$. Hence, the square root is set to zero, in these cases, placing the corresponding points at the tangent-point.

Unfortunately, KD estimates rely on the assumption of circular motion. Non-axisymmetric structures (eg. bar, spiral arms) introduce significant errors as a result of their associated streaming motions\,\cite{reid2022accuracy, ramon2018streaming, baba2009origin, xu2006distance}.

There are several key features about the KD error map of Fig. \ref{fig:intro_KE_maps} that we want to emphasise: 
\begin{itemize}
    \item There are alternating regions of positive (red) and negative (blue) errors, which are often separated by zones of low error (white). This reflects transitions between overestimated and underestimated distances. These transitions are particularly pronounced near the spiral arms and bar. 
    \item KD \added[id = anon2]{errors} are high near $\ell = 0 ^{\circ}$. This is because the denominator of Equation \ref{equ:kin_R} becomes poorly constrained with  $\sin\ell \rightarrow 0$ and $V_{\rm los} \rightarrow 0$. Consequently, regions with $\ell \lesssim 30^{\circ}$ are commonly excluded from KD-based analyses in both observational and simulation-based studies\,\cite{hunter2024testing, oort1979galactic, gomez2006errors}. Additionally, the white patches near $\ell = 0^{\circ}$ in the KD error maps of Fig.~\ref{fig:intro_KE_maps} correspond to points omitted from our analysis because they have $V_{\rm los} < = - \Theta_{0} \sin \ell$, which yields negative (unphysical) estimates for $R$.

    As a brief note, we use several cuts when applying the KD method to the simulation. We exclude some low-mass, high velocity gas parcels (mass $< 50 \; \mathrm{M_{\odot}}$ or $|V_{\rm los}| > 300$~km/s), which are occasionally generated from stellar feedback in the model. These produce unphysical KD estimates otherwise. The square‑shaped features visible in the KD map of Fig.~\ref{fig:intro_KE_maps} result from smoothing over the remaining gas after these cuts.

    \item High KD errors occur near the Sun. Analytically, $\frac{\partial R}{\partial V_{\rm los}} \propto \frac{1}{(\Theta_{0}\sin \ell + V_{\rm los})^{2}}$. Near the Sun $V_{\rm los} \approx 0$, while at larger radii $V_{\rm los}$ tends to vary between -300 and +300. This means that small variations in $V_{\rm los}$ lead to large changes in estimated $R$ close to the Sun; meanwhile at large heliocentric distances, small variations in $V_{\rm los}$ changes $R$ estimates less. 
    \item Importantly, the tangent-point circle leaves a distinct imprint on the KD error map. There are sharp transitions in KD error across the TP circle, often due to switching between positive and negative $d_{\rm k}$ solutions, corresponding to the $\pm \sqrt{R^{2} - (R_{0}\sin\ell)^{2}}$ term in Equation~\ref{equ:kin_distance} \added[id =anon]{of the Methods}. Physically, a single line-of-sight within the solar circle intersects a circular orbit at two distinct points, providing two possible distance solutions that we often alternate between across the TP circle. We also note that portions of the TP circle's circumference are outlined in white, which means that these are regions of low KD error. This indicates that some points located on the TP circle are assigned accurate distances using the KD method. 

   \item \added[id =anon]{In the bottom-right panel of Fig.~\ref{fig:intro_KE_maps},} the low KD error zones have a spiral-like structure; however, they do not consistently align with the location of gas spiral arms.
    Previous studies using 2D isothermal simulations have reported that low KD error zones trace gas density peaks along spiral arms\,\cite{hunter2024testing}. Our result is strikingly different. We attribute the difference in results to our use of 3D models. In 2D, gas is confined to the $z = 0$ plane. When gas encounters spiral-arm shocks, it cannot move vertically and is instead compressed into the gravitational potential minimum of the arms. This limits the development of non-circular streaming motions and therefore keeps KD errors low along the arms. In our 3D models, by contrast, gas can move vertically as well as in the plane; shocks thus induce stronger non‑circular motions, leading to higher KD errors along the arms. Overall, this emphasises the need for better developed models, with more advanced gas physics, in order to better constrain kinematic distances errors in the Galaxy. \added[id = anon]{
    Additionally, improved methods are required when estimating kinematic distances in regions with significant non-circular motions (eg.\,\cite{baba2026bar}). }
\end{itemize}

In this work, low KD error cuts are used to isolate gas near the simulated bar shock fronts (ie. the bar dust lanes, where gas is dense along the bars' leading edges).

\subsection*{$V_{\rm los}$ around the tangent-point circle }\label{method:TP}
\added[id = anon]{Throughout this paper, we often display $V_{\rm los}$ and $\ell$ values computed around the TP circle in our simulations. These are shown, for example, by the magenta curves of Fig.~\ref{fig:magenta_grid}, and in the top right panel of Fig.~\ref{fig:periodograms}.}

\added[id = anon]{TP $\ell$ and $V_{\rm los}$ values are calculated by projecting the low-latitude gas into the $(x,y)$ plane, identifying the TP circle, and computing mean $\ell$ and $V_{\rm los}$  values within a 0.1~kpc thick annulus centred on the TP circle (as in \EDref{fig:TP_circle}). This enables a direct comparison between TP velocities and the true terminal velocity envelope.}

\subsection*{Identifying peak features in simulated $(\ell, V_{\rm los})$ distributions}\label{method:sim_peak}

\added[id = anon2]{We have already provided a brief definition of the first and second ``peak'' features in $(\ell, V_{\rm los})$ diagrams (see \nameref{sec:second_peak} section in the main text). However, we describe our definitions in more depth here. For the first ``peak'', we begin by identifying the ridge at low longitudes corresponding to dense gas in the nuclear ring. This nuclear ring ridge is commonly identified in $(\ell, V_{\rm los})$ distributions of gas in the Milky Way; it appears as an approximately straight diagonal ridge extending from the peak $V_{\rm los}$ in Quad I to the most negative $V_{\rm los}$ in Quad IV. This structure is labelled in the CO observations of \EDref{fig:dame_image} \cite{dame2001milky}. A clear example of the nuclear ring ridge location in $(\ell, V_{\rm los})$ is also shown in \cite{fux19993d}. This ridge is generally easy to identify in $(\ell, V_{\rm los})$ because it appears very dense in molecular gas, compared to its immediate surroundings. We define the first ``peak'' as the peak $|V_{\rm los}|$ of the nuclear ring ridge (ie. the location of the most positive and most negative $V_{\rm los}$ values in this low-longitude, straight ridge). Examples of the first ``peak'' feature are clearly labelled in \EDref{fig:dame_image} for the observations and Fig.~\ref{fig:bar_shock_front} for a snapshot from Model 1 -- our MW ``surrogate'' model. For Model 2 (see \EDref{fig:Liz_sim}), the first ``peaks'' occur at relatively low $|V_{\rm los}|$ values $(\lesssim 200~\mathrm{km \; s^{-1}})$, compared to the observations or Model 1. This makes sense because Model 2 has a small bar and is not a MW ``surrogate'', but rather designed to reproduce \textsc{ngc4303} as much as possible. Once we have identified the first ``peak'', the second ``peak'' becomes easier to locate. In the observations, the second ``peak'' is obvious; we start at the first peak, then move outward to larger $|\ell|$ where a distinct cavity in the gas is present, and immediately after this the second ``peak'' appears in the terminal velocity envelope. After the second ``peak'', the terminal velocity curve declines relatively smoothly with increasing $|\ell|$. At late times in Model 1, the first and second ``peaks'' are also easy to identify because the bar is large so the two ``peaks'' are significantly separated in their longitudes. For Model 2, we begin by finding the nuclear ridge (which appears as the high density ridge at low longitudes in the right panel of \EDref{fig:Liz_sim}), and we identify the first peak to be associated with this ridge. After this, the second ``peak'' appears as the next most obvious peak in the $(\ell, V_{\rm los})$ distribution.}

\added[id = anon2]{Our definitions of first and second ``peaks'' can be used to locate bar dust lanes for other models in the literature too. For example, see the right column of Fig.~15 in \cite{englmaier1999gas}. The second ``peaks'' occur at $\ell \approx \pm 4^{\circ}$. These lines-of-longitude intersect the TP circle in the bar region, consistent with our argument. Unfortunately, the gas distribution from \cite{englmaier1999gas} does not feature clear dust lanes, making it unideal for testing our hypothesis. Alternatively, see the first row of Fig.~8 in \cite{rodriguez2008gas}. The second ``peaks'' occur at $\ell \approx \pm 11^{\circ}$, which are the approximate longitudes where the TP circle intersects the bar dust lanes. The best counterexample we could find to our argument is the hydrodynamical model displayed in the left column of Fig.~15 from \cite{fux19993d}. This simulated $(\ell, V_{\rm los})$ distribution only features a clear second ``peak'' in Quad IV at $\ell \approx -15^{\circ}$, which maps back to the intersection of the TP circle with the 3-kpc arm in $(x,y)$. While the 3-kpc arm is not a bar dust lane, we emphasise that it is not a spiral arm either. Rather the 3-kpc is a bar associated structure; it is a spiral-like feature that surrounds the bar and probably follows $X_1$ orbits \cite{kumar2025expanding}. Hence, even in this case the second ``peak'' can still be used to study the bar. We note that many previously simulated $(\ell, V_{\rm los})$ distributions in the literature do not have a clear second ``peak'' at all. In some cases this may result from the colormap chosen, along with the decision to combine different phases of gas when constructing these $(\ell, V_{\rm los})$ distributions (eg. \cite{li2022gas}). In other cases, the absence of a second ``peak'' makes the model inconsistent with observations of gas in the Milky Way.}

\added[id = anon]{In Fig.~\ref{fig:magenta_grid}, we label the ``peak'' features of the simulated $(\ell, V_{\rm los})$ distributions with the blue scatter points. Additionally, in panel (d) of Fig.~\ref{fig:periodograms}, we display the longitude coordinate of these ``peak'' features over the course of the simulation.} \added[id = anon]{These were identified by first generating an $(\ell, V_{\rm los})$ distribution for the relevant snapshot, tracing the terminal velocity curve (the maximum \(|V_{\rm los}|\) in each longitude bin), removing outliers, smoothing the resulting profile, and locating peaks along it using \texttt{scipy.signal.find\_peaks}. The identified peak positions were overlaid on the \((\ell, V_{\rm los})\) diagrams to check their accuracy using visual inspection,
with some results adjusted manually where required.}

\subsection*{Methods for locating the Bar Dust lanes}\label{sec:test}

\added[id= anon]{In simulations, we know that a correlation between $(\ell, V_{\rm los})$ ``second peaks'' and the bar dust lanes commonly exists. Ultimately, we want to use the ``second peak'' features in observed CO and HI $(\ell, V_{\rm los})$ diagrams to predict the location of the Galactic bar dust lanes. We begin by pracitising our method on a simulation snapshot.}

\added[id = anon]{We take an $(\ell, V_{\rm los})$ diagram for a representative snapshot in our model, identify the second peak (at $\ell >0$), and project its coordinates back into $(x,y)$ space to check whether it coincides with the TP–bar dust lane intersection.}\added[id = anon]{This projection can be performed using two similar methods:}
\begin{itemize} 
    \item Longitude-TP intersection. Determine the longitude of the ``second peak'', draw the corresponding line of longitude in $(x,y)$ space, and compute its intersection with the tangent-point circle. If the terminal curve changes sharply where the TP circle meets the dust lane, this intersection should coincide with the secondary peak. 
    \item The Kinematic Distance Method. Substitute the $(\ell, V_{\rm los})$ \added[id = anon]{coordinates} of the second peak into the kinematic distance equations \added[id = anon]{(see Methods, Equations~\ref{equ:kin_R} and \ref{equ:kin_distance})} and hence estimate the gas parcel's position in $(x,y)$ space. \added[id = anon2]{For most barred snapshots in Model 1, a low KD error zone $(< 2 \%)$ appears in the gas immediately adjacent to the bar shock fronts (eg. see Fig.~\ref{fig:bar_shock_front}). As a result, KD methods are often highly accurate at locating the bar shock fronts in this model.}
\end{itemize}

In Supplementary Fig.~\ref{fig:second_peak}, we apply both methods to the $t = 3.95$~Gyr snapshot. The right panel marks the first and second $(\ell, V_{\rm los})$ peaks with cyan and lime green points, while the left panel shows their $(x,y)$ positions computed with the kinematic distance method. In the $(x,y)$ panel, we also overlay the TP circle (white dashed) and lines of longitude for both peaks (black dashed). The lime green points lie exactly at the TP circle--bar dust lane intersection, confirming that the second peak in $(\ell, V_{\rm los})$ maps to this location at $t = 3.95$~Gyr in the model. Both described methods accurately recover this intersection point. 

\added[id = anon]{Additionally, we note that the first and second ``peaks'' enclose an obvious gas poor region in the CO $(\ell, V_{\rm los})$ observations (\EDref{fig:dame_image}). This is highly reminiscent of the gas-poor ``shock cavity'' in the bar region between the dust lanes and nuclear ring. If this gas-poor region in $(\ell, V_{\rm los})$ does map to the bar ``shock cavity'', it would make sense that the bar dust lanes lie close to the peaks.}

\subsection*{Measuring bar length}\label{method:bar_length}

The cusped $x_1$ orbit should approximately follow the bar's shock fronts, or dust lanes, and encase the gas-poor ``shock cavity''.\,\cite{binney1991understanding, morris1996galactic, gerhard1996dynamics} (described in Introduction; see\,\cite{downes1996co}, Fig.~10 for a diagram). \added[id = anon]{In this work, we measure the bar semi-major axis radius, $a_{\rm bar}$, using the length of the cusped $x_1$ orbit. In simple terms, we assume that the Norma masers approximately trace the cusped $x_1$ orbit, and hence we are able to estimate the length of this orbit's semi-major and semi-minor axes. In Fig.~\ref{fig:MW_bar_length}, we draw cyan dashed lines to mark the location where we believe the cusped $x_1$ orbit to be. This is done by eye. Hence, by measuring the length of these cyan dashed lines, we determine estimates for $a_{\rm bar}$ and $b_{\rm bar}$ (quoted in Sec.~\nameref{sec:bar_properties}). }

\added[id = anon]{Many different methods are used in the liteature to measure bar length. It is well-known that each method often yields different results for the same intrinsic bar. Other commonly used methods include Fourier decomposition (e.g.,\,\cite{ wegg2015structure, hilmi2020fluctuations, dehnen2023measuring}), ellipse fitting\,\cite{wegg2015structure, athanassoula2002morphology}, and overdensity contour mapping\,\cite{hilmi2020fluctuations}. We apply several of these methods to our simulation snapshot in order to compare their different estimates for bar length. Firstly, we estimate the radius at which the stellar overdensity profile along the bar's semi-major axis decreases to $50 \%$ from its maximum $(L_{\mathrm{cont}})$\,\cite{hilmi2020fluctuations}. Secondly, we compare the stellar density profiles along the bar's semi-major and semi-minor axes. Their difference is zero at the bar's centre, then increases to a maximum, and finally decreases as we move into the surrounding disc. The radius, outward from the maximum, where the difference falls below $30 \%$ of the semi-major axis density gives an estimate for bar length $(L_{\mathrm{prof}})$\,\cite{wegg2015structure, hilmi2020fluctuations, athanassoula2002morphology, scannapieco2012bars}. Thirdly, bar length is estimated using the radius at which the $m = 2$ Fourier amplitude drops to $70 \%$ of its peak ($L_{m =2}$)\,\cite{ghosh2024looking}.}

\added[id = anon]{The original simulation snapshot at $t = 3.95$~Gyr (\textit{before} scaling and rotation to match the Norma masers) gives: $L_{\rm prof} = 4.3$~kpc, $L_{\mathrm{m = 2}} = 3.2$~kpc, $L_{\mathrm{cont}} = 3.95$~kpc and $L_{\mathrm{x1}} = 3.4$~kpc. This results in a mean and standard error of the mean (sem): $L_{\rm bar} = 3.7 \pm 0.5$~kpc.  In comparison, the simulation snapshot that \textit{is} scaled and rotated to match the Norma masers (see Fig.~\ref{fig:bar_shock_front}) gives: $L_{\rm prof} = 4.95$~kpc, $L_{\mathrm{m = 2}} = 3.9$~kpc, $L_{\mathrm{cont}} = 4.65$~kpc and $L_{\mathrm{x1}} = 4.3$~kpc. Here, the mean is: $L_{\rm bar} = 4.5 \pm 0.5$~kpc.  We note that scaling the model, unevenly in x and y (by a factor of 1.16 in x  and 1.24 in y), without changing the mass, makes the model less realistic. A larger bar would likely have more mass and a different density profile to our scaled version. Since some of these bar length measurements depend on the density distribution within the bar, uncertainties may arise from the nature of the scaling.}

\added[id = anon]{
Final quoted uncertainties for bar semi-major radius ($a_{\rm bar}$), semi-minor radius $(b_{\rm bar})$ and orientation angle $(\phi)$ (quoted in Abstract and Sec.~\nameref{sec:bar_properties}) are found using the parallax uncertainties for the Norma masers\,\cite{reid2019trigonometric}. Parallax uncertainties are propagated into x and y coordinate space, and the result is visualised with the uncertainty bars in Fig.~\ref{fig:MW_bar_length}. Then, the simulation snapshot is both scaled and rotated by eye such that we find the most extreme cases where the bar shock front lies within the uncertainty bars of the Norma masers (ie. smallest/largest bar size, highest/lowest inclination angle values). Finally, maximum and minimum possible $a_{\rm bar}$, $b_{\rm bar}$ and $\phi$ values are estimated using the hypothesised location of the $x_1$ orbit in the scaled and rotated simulation. We note that the final quoted uncertainties do not include variation introduced from different bar length methods. Many of the previously mentioned bar length methods rely more strongly on stellar density profiles in the bar, while our approach of estimating the $x_1$ orbit size is more strongly constrained by the bar dust lane location and associated gas distribution. }

\clearpage

\bmhead{Data Availability}
Our Code Ocean capsule contains many pre-computed arrays, generated through analysis of the simulations. These saved arrays should be sufficient for reproducing any of the main figures in this work. Correspondence and requests for further materials should be addressed to Hillary Davis (hillary.davis@sydney.edu.au). For access to the Milky Way ``surrogate'' simulation, requests should be addressed to Thor Tepper-Garc{\'\i}a (thorsten.teppergarcia@sydney.edu.au).

\bmhead{Code Availability}
Our Code Ocean capsule contains all code and pre-computed arrays for reproducing the main figures of this paper.

\bmhead{Acknowledgements}
\added[id = anon]{We thank the anonymous reviewers for their suggestions, which we believe significantly improved the paper. We also thank Robin Joshi for his help with setting up the second model.  ChatGPT (OpenAI) was used for minor editing of code and writing, and for figure-presentation suggestions. All suggestions were reviewed and verified by the authors, who take full responsibility for the final content.}

\bmhead{Funding}
TTG acknowledges financial support from the Australian Research Council (ARC) through an Australian Laureate Fellowship awarded to JBH. The computations and data storage were enabled by two facilities: (i) the National Computing Infrastructure (NCI) Adapter Scheme, provided by NCI Australia, an NCRIS capability supported by the Australian Government; and (ii) LUNARC, the Centre for Scientific and Technical Computing at Lund University (resource allocations LU 2023/2-39 and LU 2023/12-6). EJI acknowledges the support of the Australian Research Council through Discovery Project DP220103384 (CIs:\,JBH,\,Ken Freeman). OA acknowledges support from the Knut and Alice Wallenberg Foundation, the Swedish Research Council (grant 2025-04892), the Swedish National Space Agency (SNSA grants 2023-00164 and 2025-00405), the LMK foundation, and eSSENCE, a Swedish strategic research programme in e-Science.

\bmhead{Author contribution}

H.D. planned the project, carried out all the analysis, generated the figures and led the writing of the manuscript. J.B.H supervised the project, engaged in extensive discussion, contributed to the interpretation of results and edited the manuscript. T.T.G. made the MW ``surrogate'' simulation (Model 1) -- the main model we focus on in this work, and he also wrote the N-body/hydrodynamic simulation section in the Methods, and engaged in discussion of results. O.A. worked with T.T.G to make the MW ``surrogate'' model, and provided extensive feedback on both the project and the manuscript. E.J.I. made Model 2, engaged in discussion about the simulation, helped with writing its description in the Methods section, and edited the manuscript.

\bmhead{Competing interests}
The authors declare no competing interests.

\clearpage

\clearpage

\section*{Tables}

\noindent

\setcounter{figure}{0}
\renewcommand{\thefigure}{\arabic{figure}}
\renewcommand{\figurename}{Extended Data Fig.}

\begin{table}[!ht]
    \centering
    \caption{The coordinates of the second and third peaks in the CO $(\ell, V_{\rm los})$ observations of\,\cite{dame2001milky} (their Figure 3). We also include the corresponding $(x_{k}, y_{k})$ positions of gas estimated using the kinematic distance method (found with $R_{0} = 8.15$\,kpc and $\Theta_{0} = 236$\,km\,s$^{-1}$). In Quadrant~I, only the first and second peaks are present. Quadrant~IV also has a third peak, but this is not the focus of the present work.}
    \label{tab:CO_peak_estimates}
    \begin{tabular}{l c c c c c}
        \hline
        Peak & $\ell \, (\mathrm{deg})$ & $V_{\rm los}$ (km s$^{-1}$) & $d_{\rm k}$ (kpc) & $x_{\rm k}$ (kpc) & $y_{\rm k}$ (kpc) \\
        \hline
        Second peak Quad I  & $22^{\circ}$    & $124$   & 7.56 & -1.14 &  2.83 \\
        Second peak Quad IV & $-16.5^{\circ}$ & $-150$  & 7.81 & -0.66 & -2.22 \\
        Third peak Quad IV  & $-28^{\circ}$   & $-111$  & 7.20 & -1.79 & -3.38 \\
        \hline
    \end{tabular}
\end{table}

\begin{figure}
    \includegraphics[width=1\textwidth]
    {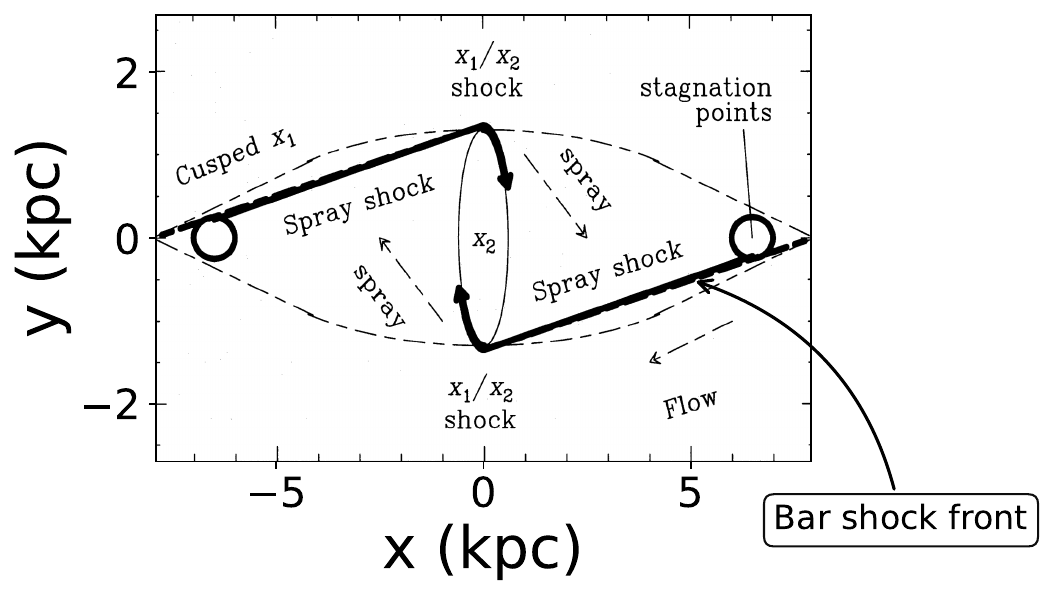}
    \caption{\added[id = anon]{\textbf{A schematic of $x_1$ and $x_2$ orbits in the bar region.} This schematic has been taken from\,\cite{downes1996co}, with the bar shock front labelled by us, $\copyright$AAS.. It shows the theorised location of $x_2$ orbits and the cusped $x_1$ orbit in the bar region. \added[id = anon3]{The dashed black curves, which mark out an approximate-ellipse shape with pointed ends, show the location of the cusped $x_1$ orbit. The solid black oval at shorter radii is the location of an $x_2$ orbit.} Shocks occur near the cusped $x_1$ orbit, convert atomic gas into the molecular form, and cause gas to either fall inward to the $x_2$ family or spray across to the opposite side of the bar.}} 
    \label{fig:bar_spiral_schematic}
\end{figure}

\begin{figure}
    \centering
    \includegraphics[width=\textwidth]
    {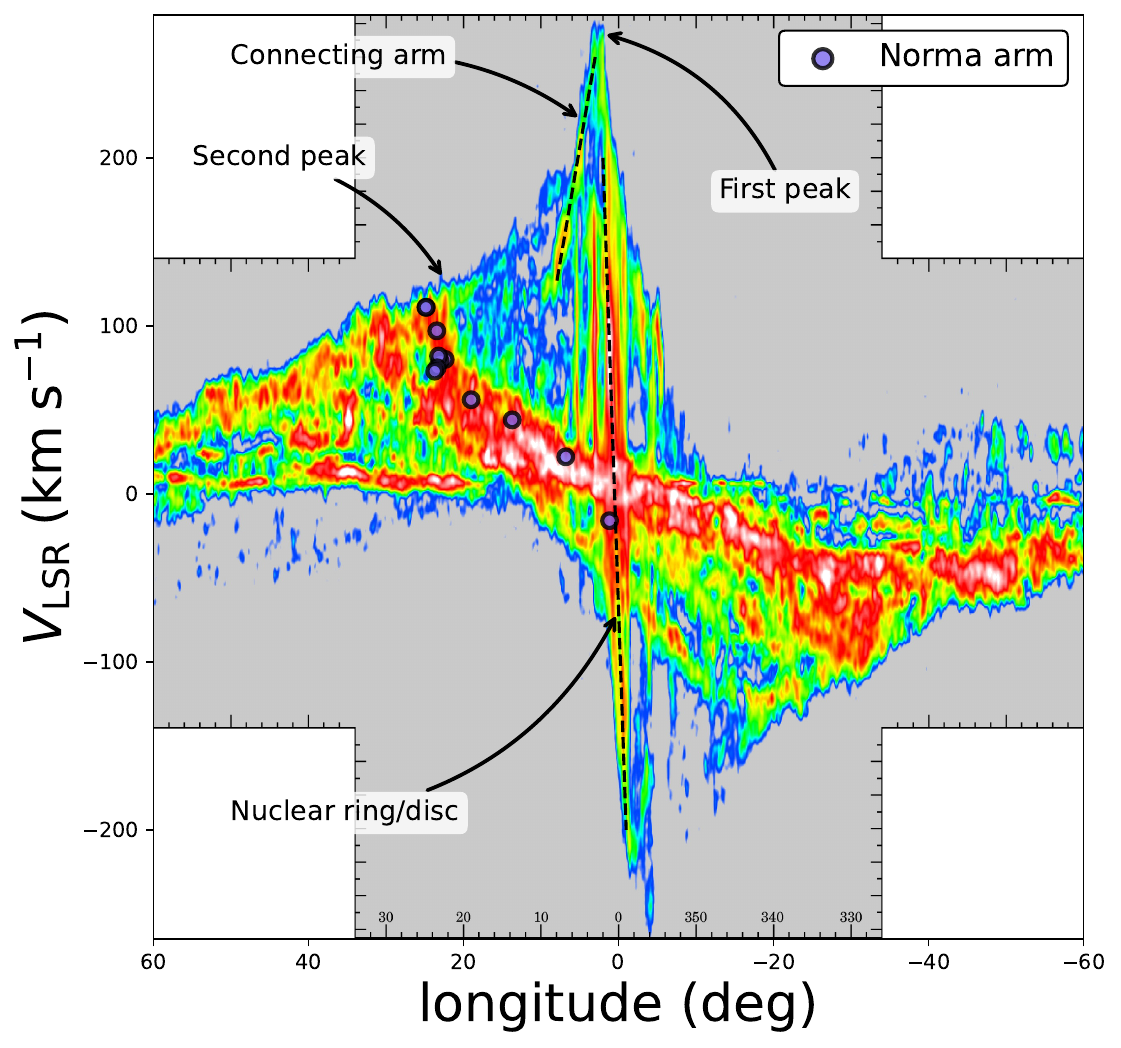}
    \caption{\added[id = anon]{\textbf{Observations of CO gas in the Milky Way.} $(\ell, V_{\rm los})$ observations of CO gas in the Milky Way, adapted from\,\cite{dame2001milky}, with permission. On the y-axis, we use velocity relative to the local standard of rest, $V_{\rm LSR}$, since this is the conventional term used in \cite{dame2001milky}. However, in this work, $V_{\rm LSR}$ is equivalent to $V_{\rm los}$, and defined in the same way. We overlay black dashed lines to mark the hypothesised location of the nuclear ring and the Quad I connecting arm in the literature\,\cite{fux19993d, reid2016parallax}. We label, what we have termed the ``first peak'' and ``second peak'' features in the $(\ell, V_{\rm los})$ observations. These are locations in the terminal curve of the underlying $(\ell, V_{\rm los})$ distribution, where there is a distinct peak in $V_{\rm los}$. Finally, we overlay the location of the Norma arm masers from\,\cite{reid2019trigonometric} to highlight that this is currently believed to be the location of a spiral arm, but we argue that this is the Quad I bar shock front.} \added[id = anon2]{The colour scale shows the logarithm of CO brightness temperature integrated over the central $4^{\circ}$ of Galactic latitude ($\mathrm{log} \int T_{\rm mb} db$), in units of K arcdeg. Red represents the brightest CO emission and dark blue the dimmest \cite{dame2001milky}. \added[id = anon3]{Please refer to \cite{dame2001milky} for a full colormap.}}}
    \label{fig:dame_image}
\end{figure}

\begin{figure}
    \centering
    \includegraphics[width=1.1\textwidth]
    {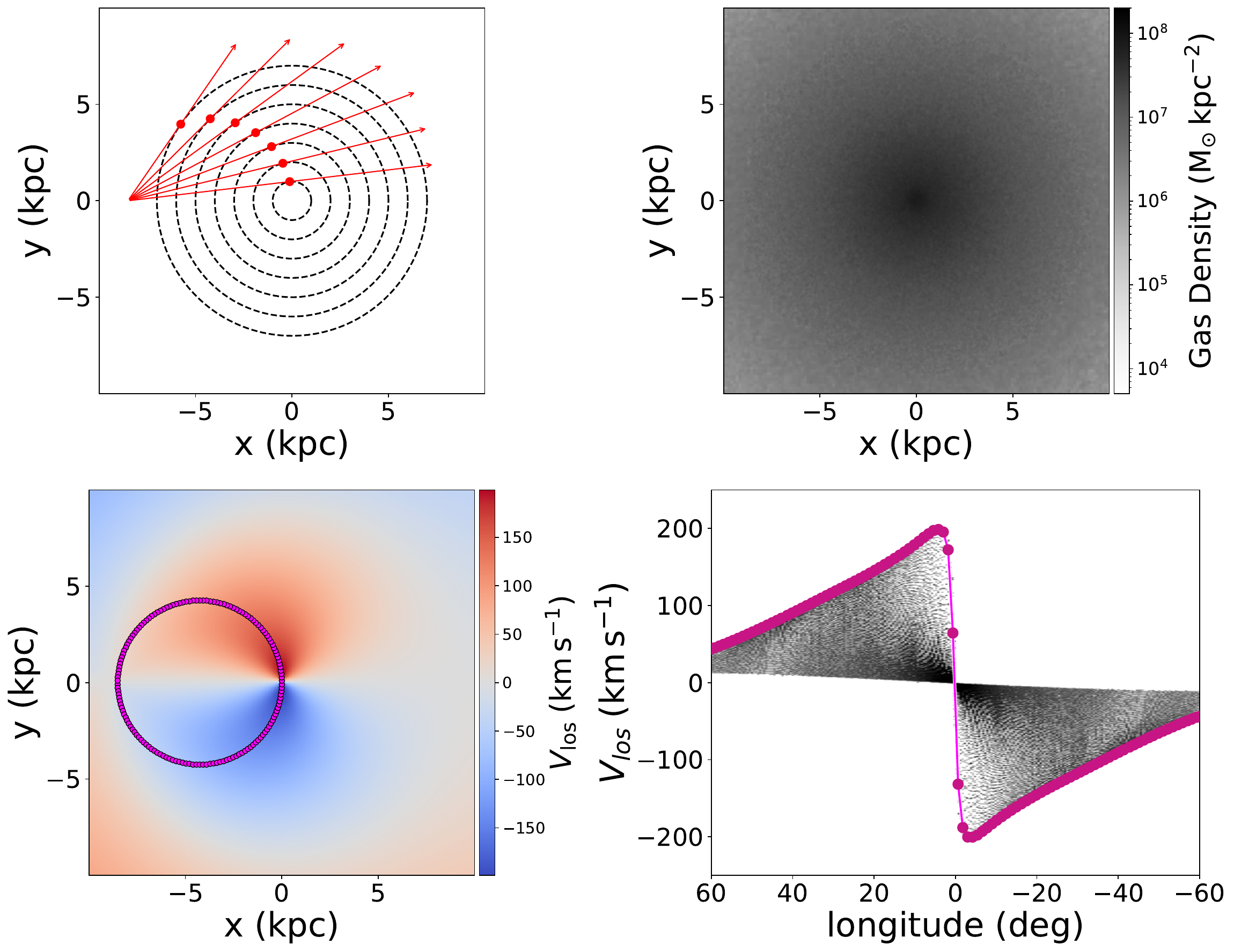}
    \caption{\added[id = anon]{\textbf{The tangent-point circle and terminal velocities at the beginning of our model. } Top left panel: This schematic shows the tangent points for pure circular motion, adapted from\,\cite{davis2025hi}. The black dashed circles represent gas on circular orbits at different radii. Lines-of-sight (red) are tangent to these orbits at a single point, marked by the red points. These red tangent points trace a circle between the Sun at $(x,y) = (-8.5, 0)$ and the Galactic centre, which we refer to as the \textit{tangent-point circle}. Top right panel: Gas $(x,y)$ density projection for a low-latitude slice ($|b| < 2^{\circ}$) at $t = 0$~Gyr in our MW “surrogate” simulation. Bottom left panel: Line-of-sight velocity, $V_{\rm los}$, for the same slice, projected into the $(x,y)$ plane. Along the TP circle, we compute the mean $\ell$ and $V_{\rm los}$ at each of the magenta points. Bottom right panel: Gas distribution in $(\ell, V_{\rm los})$ space at $t = 0$~Gyr. The magenta points indicate the corresponding $\ell$ and $V_{\rm los}$ values derived from the TP circle in the bottom left panel.} }
    \label{fig:TP_circle}
\end{figure}

\begin{figure}
    \hspace{-1.8cm}
    \includegraphics[width=1.3\textwidth]
    {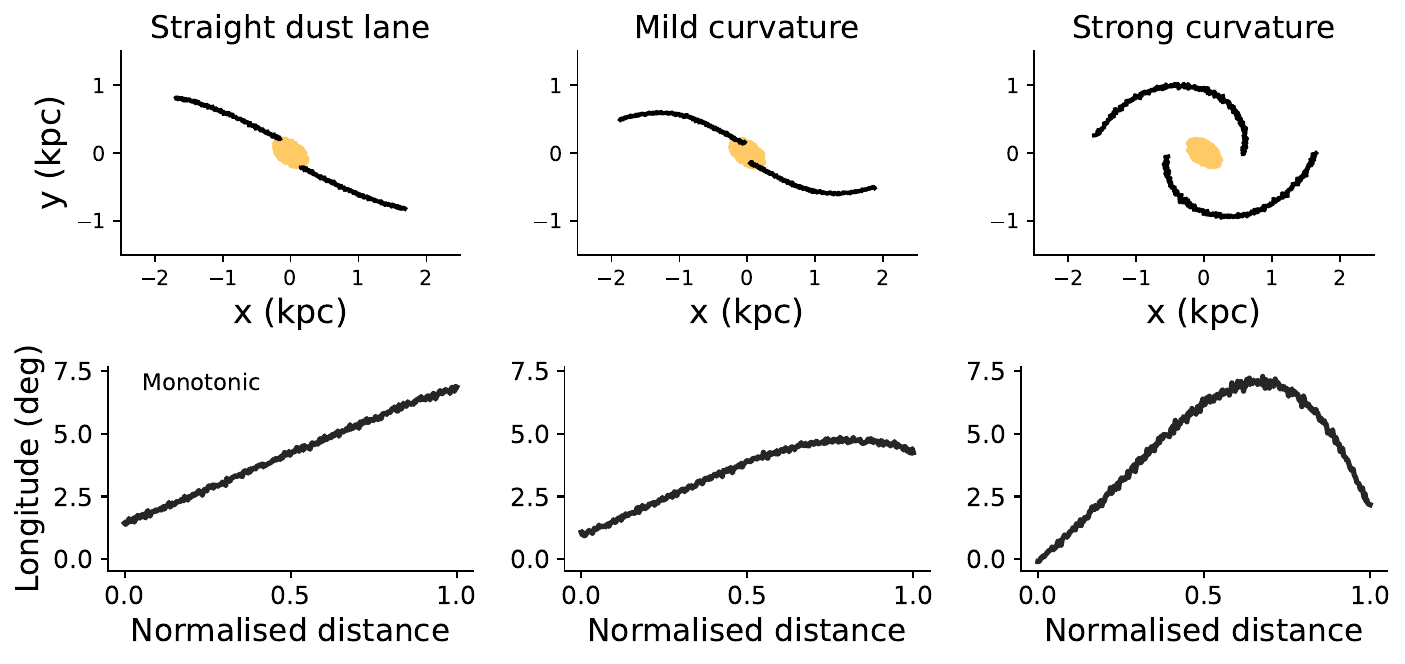}
    \caption{\added[id = anon]{\textbf{Reproducing a straight connecting arm ridge in $(\ell, V_{\rm los})$ space.} Demonstration that we need a straight dust lane to reproduce a straight ridge that decreases monotonically with longitude in $(\ell, V_{\rm los})$ space. The top row shows a simple schematic of bar dust lanes with various curvatures in $(x,y)$. \added[id = anon3]{The yellow ovals at the centre represent the nuclear ring, and the black curves represent dust lanes which we expect to follow the leading edges of bars with various morphologies.} The bottom row presents the corresponding longitude values for the Quad I dust lane. For strongly curved dust lanes (right panel), longitude initially increases but subsequently decreases as the lane wraps around the bar. In contrast, for a straight dust lane, we recover a straight line of continuously increasing longitude values. In this schematic, the bar is at approximation $35^{\circ}$ inclination and the sun at $(-8.5, 0)$~kpc. Increasing the bar angle can reduce the extent of the turnover in longitude for curved dust lanes, but does not eliminate it. }}
    \label{fig:dust_lane_shape}
\end{figure}

\begin{figure}
    \centering
    \includegraphics[width=1.1\textwidth]
    {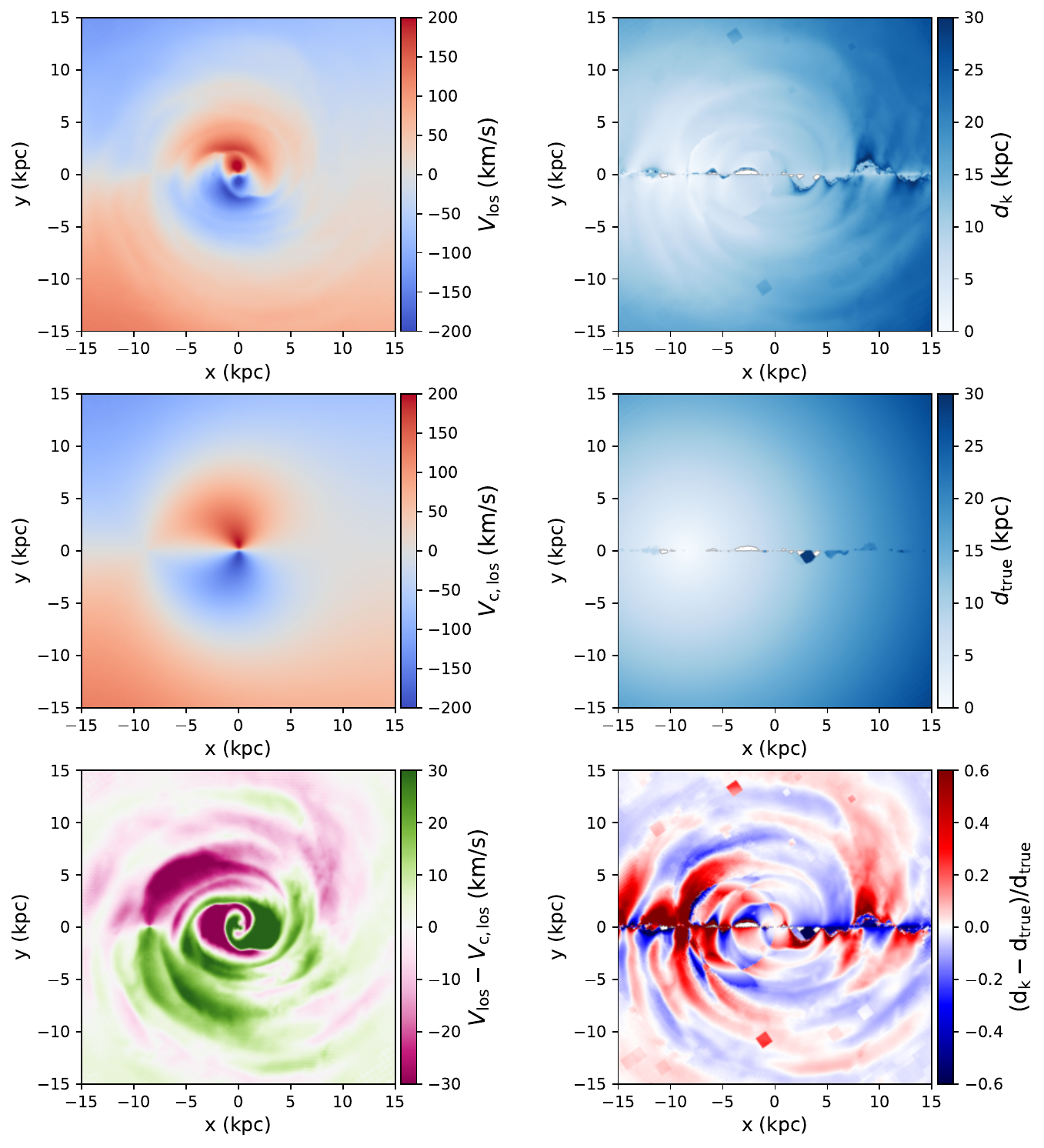}
    \caption{\textbf{Non-circular motions and kinematic error maps in the simulation.}  This is a reproduction of Figure 10 from\,\cite{hunter2024testing}. The first column presents, from top to bottom: mass-weighted $V_{\rm los}$ projected along the z-axis, the corresponding $V_{\rm los}$ projection for the case of pure circular motion in an azimuthally-averaged axisymmetric potential, and the difference $V_{\rm los} - V_{\rm c,los}$. The second column shows, from top to bottom: distances estimated with the KD method $d_{\rm k}$, true distances in the simulation $d_{\rm true}$, and the resulting KD error.}
    \label{fig:hunter_image}
\end{figure}

\begin{figure}
    \centering
    \includegraphics[width=\textwidth]
    {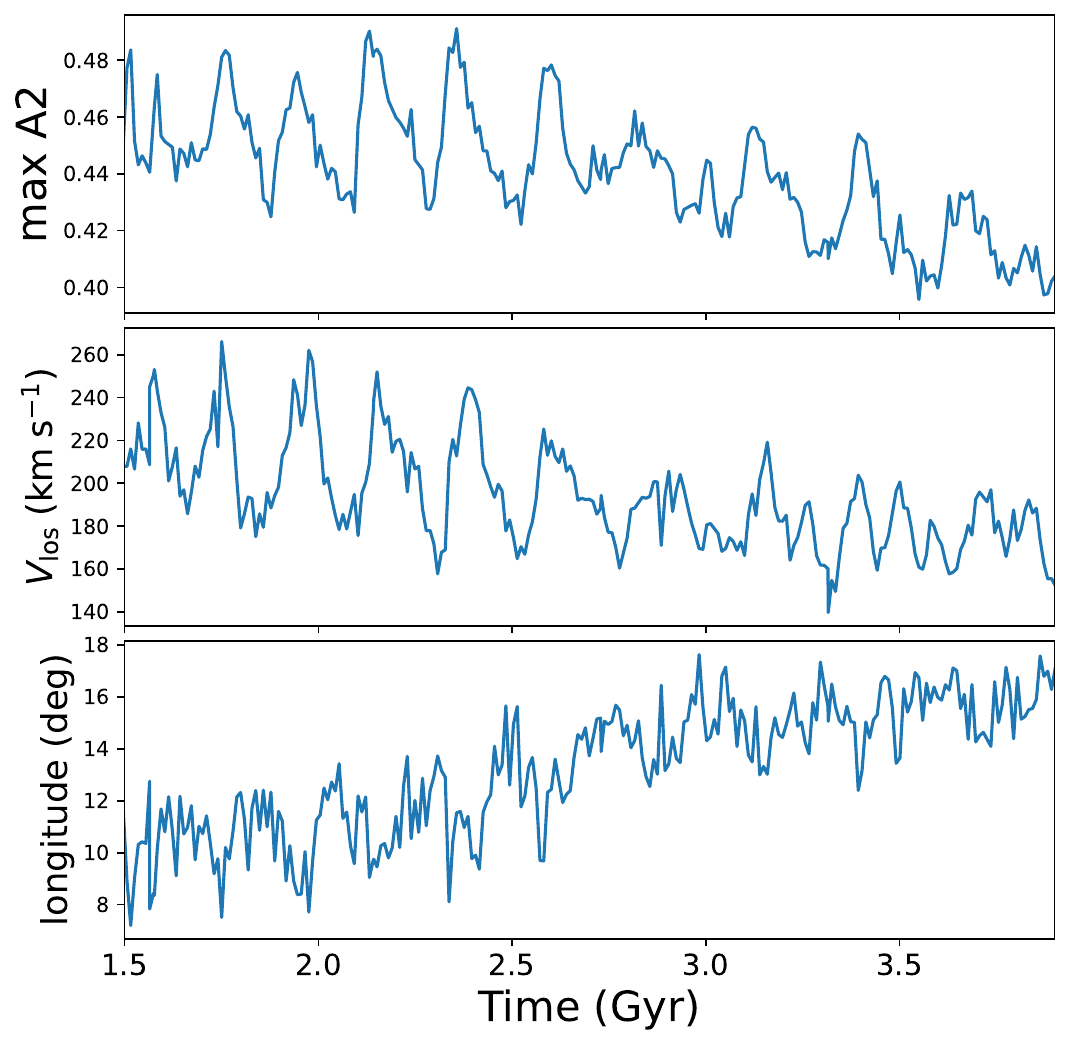}
    \caption{\added[id = anon]{\textbf{Second peak velocity and bar-spiral arm attachment.} Top row: Maximum $A_2$ values over time in our Milky Way ``surrogate'' model. Middle row: $V_{\rm los}$ values of the Quad I ``second peak'' feature oscillating over time in the simulation. These strongly correlate with the $A_2$ values. Bottom row: Longitude values of the Quad I ``second peak'' feature evolving over time in the model.  }}
    \label{fig:second_peak_fluctuations}
\end{figure}

\begin{figure}
    \hspace*{-0.7cm}
    \includegraphics[width=1.2\textwidth]
    {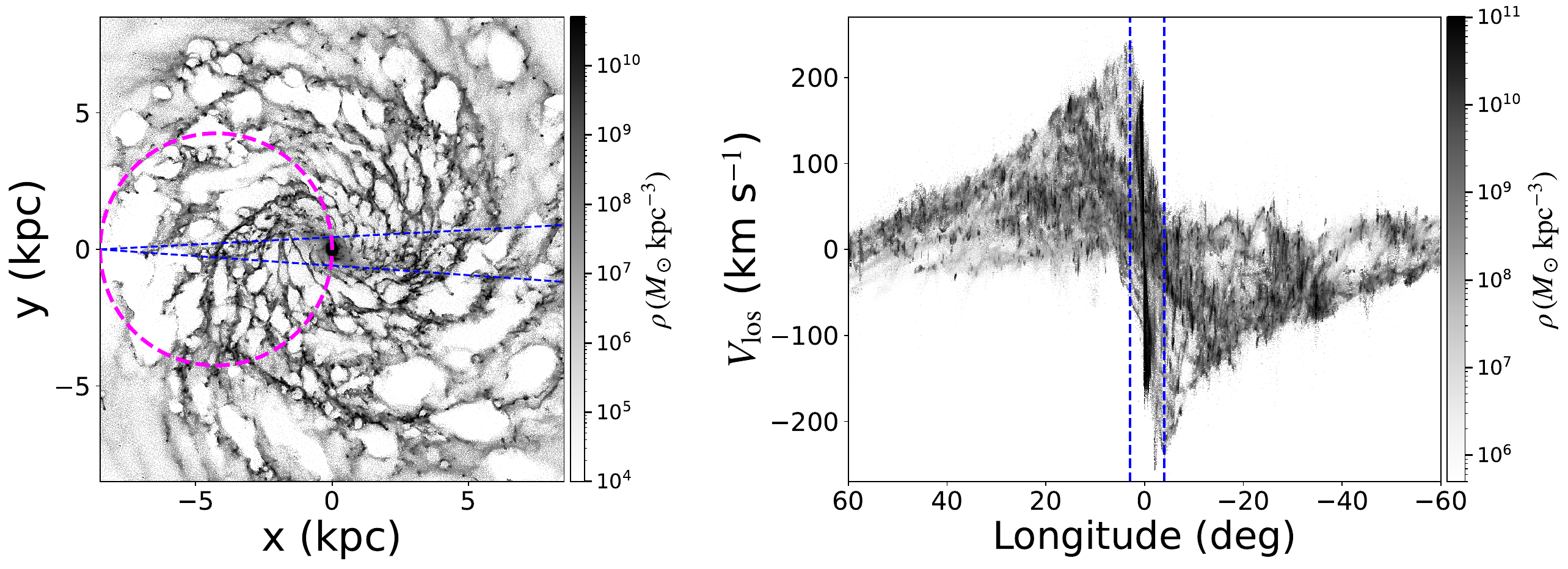}
    \caption{\added[id = anon]{\textbf{Locating the bar dust lanes in a second galaxy simulation.} $(x,y)$ gas density distribution and corresponding $(\ell, V_{\rm los})$ distribution for $t = 920$~Myr in Model 2. The bar is oriented at $25^{\circ}$ to the Sun–Galactic centre line. Dark blue dashed lines mark locations of the ``second peaks''. These occur at $\ell = 3^{\circ}$ and $\ell = -4^{\circ}$. We overlay the corresponding lines-of-longitude in dark blue on the $(x,y)$ map; the magenta circle marks the TP circle. We see that the bar dust lanes in $(x,y)$ occur at the approximate intersection of the ``second peak'' lines-of-longitude with the TP circle. This is consistent with the result we found in our MW ``surrogate'' simulation.} \added[id = anon2]{For a clear definition of first and second ``peaks'' in $(\ell, V_{\rm los})$ diagrams, please refer to Sec.~\ref{method:sim_peak} of the Methods.}}
    \label{fig:Liz_sim}
\end{figure}

\clearpage
\setcounter{figure}{0}
\setcounter{table}{0}

\renewcommand{\tablename}{Supplementary Table}
\renewcommand{\figurename}{Supplementary Fig.}

\section*{Supplementary Information }

\begin{table*}[!ht]
\centering
\caption{Galaxy model parameters. Columns 1-6 list the galactic component name, its initial functional form, total mass, radial scale length, cut-off radius, and particle number, respectively.}
\label{tab:s3b}
\resizebox{\textwidth}{!}{%
\begin{tabular}{llccccc}
\hline
Component & Profile & Total mass & Radial scalelength & Cut-off radius & Particle count \\
 &  & ($10^{10} \mathrm{M_{\odot}}$) & (kpc) & (kpc) & ($10^6$) \\
\hline
DM halo       & NFW                & 118   & 19   & 250 & 1   \\
Stellar bulge & Hernquist          & 1.25  & 0.6  & 2.0 & 0.1 \\
Stellar disc  & Exp, $\mathrm{sech}^2$ & 4.31  & 2.5  & 25  & 1   \\
Gas disc      & Exp, $\mathrm{sech}^2$ & 0.46  & 3.5  & --  & 2   \\
\hline
\end{tabular}%
}

\end{table*}

\begin{figure}[!ht]
    \centering
    \hspace{-1.4cm}
    \includegraphics[width=1.1\textwidth]
    {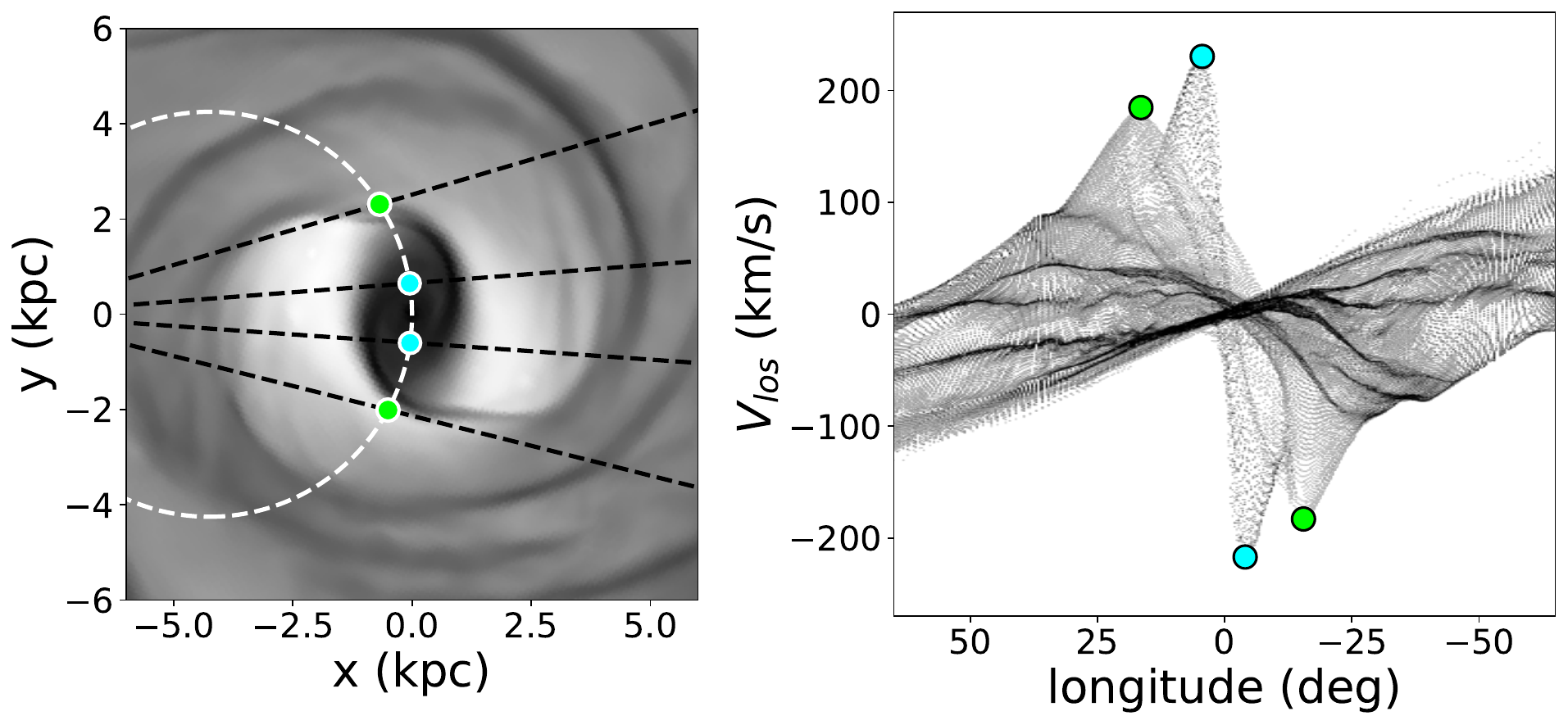}
    \caption{\textbf{Using first and second peak longitudes to constrain the bar.}  Left panel shows $(x,y)$ gas density distribution for $t = 3.95$~Gyr snapshot. Right panel is the corresponding $(\ell, V_{\rm los})$ distribution. Cyan and lime green points mark the location of the first and secondary $(\ell, V_{\rm los})$ peaks. We use the kinematic distance method to estimate the $(x,y)$ location corresponding to these peaks, with the result presented using the lime and cyan markers in the left panel. In $(x,y)$, the tangent-point circle (white dashed) and lines-of-longitude corresponding to the peaks (black dashed) are also overlaid.}
    \label{fig:second_peak}
\end{figure}

\end{document}